\documentclass[npg]{copernicus}
\usepackage{natbib}
\usepackage{multirow}
\usepackage{pifont}
\usepackage{color}
\usepackage{layout}

\def\mathi{\mathrm{i}}

\title{Resurrecting Dead-water Phenomenon}
\author[1,*]{Matthieu J. Mercier}
\author[1,**]{Romain Vasseur}
\author[1]{Thierry Dauxois}
\affil{Universit\'e de Lyon, Laboratoire de Physique de l'\'Ecole Normale Sup\'erieure de Lyon, CNRS, France.}
\affil[*]{now at: Department of Mechanical Engineering, MIT, Cambridge, MA 01239 USA.}
\affil[**]{now at: Institut de Physique Th\'eorique, CEA Saclay, 91191 Gif Sur Yvette, France\\ and LPTENS, Ecole Normale Sup\'erieure, 24 rue Lhomond, 75231 Paris, France.}
\date{\today}
\correspondence{mmercier@mit.edu, Thierry.Dauxois@ens-lyon.fr}

\begin{document}

\maketitle

\begin{abstract}
We revisit experimental studies performed by Ekman on dead-water~\citep{bib:Ekman1904} using modern techniques in order to present new insights on this peculiar phenomenon. We extend its description to more general situations such as a three-layer fluid or a linearly stratified fluid in presence of a pycnocline, showing the robustness of dead-water phenomenon. We observe large amplitude nonlinear internal waves which are coupled to the boat dynamics, and we emphasize that the modeling of the wave-induced drag requires more analysis, taking into account nonlinear effects.
\end{abstract}

%% VOCABULARY %%
% proue (avant) = bow
% poupe (arrière) = stern
% ligne de flottaison = waterline
% coque = hull

%%%%%%%%%%%%%%%%%%%%%%%%%%%%%%%%%%%%%%%%%%%%%%%%
%%%%%%%%%%%%%% INTRODUCTION %%%%%%%%%%%%%%%%%%%%
\bigskip
\emph{Dedicated to Fridtj\"of Nansen born 150 years ago (10 October 1861).}

\section{Introduction}
For sailors, the dead-water phenomenon is a well-known peculiar phenomenon, when a boat evolving on a two-layer fluid feels an extra drag due to waves being generated at the interface between the two layers whereas the free surface remains still. Interestingly, one finds reports of similar phenomena in the Latin literature when Tacitus described a flat sea on which one could not row a boat, North of Scotland and of Germany,  in the Agricola~\citep{bib:Tacitus1} and in the Germania~\citep{bib:Tacitus2}.

This effect is only observed when the upper part of the fluid is composed of layers of different densities, due to variations in salt concentration or temperature. An important loss of steering power and speed is experienced by the boat, which can even undergo an oscillatory motion when the motors are stopped.

In this paper, we present detailed experimental results on the dead-water phenomenon as shown in the video by~\cite{bib:videoSanAntonio}.  The material is organized as follows. In the remaining of this section,
we briefly review the different studies of this phenomenon, either directly related to Ekman's work or only partially connected to it.
Section~\ref{experimentalsetup} presents the experimental set-up. The case of a two-layer fluid is addressed in Sec.~\ref{twolayer}, followed by the case
with a three-layer fluid in Sec.~\ref{threelayer}. The more realistic stratification with a pycnocline above a linearly stratified fluid is finally discussed in Sec.~\ref{pycnocline}.
Our conclusions, and suggestions for future work are presented in section~\ref{sct:conclusion}.

\subsection{Ekman's PhD Thesis}
\label{sec:EkmanPhD}

V. W. Ekman was the first researcher to study in detail the origin of the dead-water phenomenon. His work as a PhD student~\citep{bib:Ekman1904} was motivated by the well-documented report from the Norvegian Artic explorer Fridtj\"of Nansen who experienced it while sailing on the {\it Fram} near ``Nordeski\"{o}ld" islands in 1893~\citep{bib:Nansen1897}.

%% Ekman thesis results + Added results from other works %%
Several aspects of the phenomenon have been described by Ekman, who did experiments with different types of boat evolving on a two-layer fluid. We note $\rho_1$ and $h_1$ the density and depth of the upper layer, and $\rho_2$ and $h_2$ those of the lower layer.

% Drag vs speed %
i) First of all, the drag experienced by the boat evolving on the stratified fluid is much stronger than the one associated with an homogeneous fluid. This difference is due to wave generation at the interface between the two layers of fluid, pumping energy from the boat.
\begin{figure}
\begin{center}
\begin{picture}(8.5,3.75)
\put(0.25,-1){\includegraphics[width=8\unitlength]{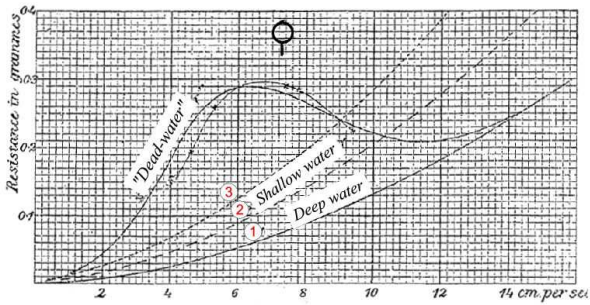}}
\end{picture}
\end{center}
\caption{Drag-speed relations given by Ekman for a boat dragged on a layer of fresh water ($\rho_1=1.000$~g~cm$^3$) of depth $h_1=2.0$~cm, resting above a salt layer ($\rho_2=1.030$~g~cm$^3$), and compared to the homogeneous case ``deep water" with a water level of \ding{192}~$23$~cm, and ``shallow water" corresponding to a smaller depth \ding{193}~5~cm, and \ding{194}~2.5~cm (taken from~\cite[Fig.~8, Pl.~VI][]{bib:Ekman1904}). Experimental points are crosses and continuous lines are models.}\label{fig:Fv_Ekman}
\end{figure}
This effect is the strongest when the boat's speed is smaller than the maximum wave speed~\citep{bib:Gill1982}, defined as
\begin{equation}
c_{\phi}^{m}=\sqrt{g\frac{\rho_2-\rho_1}{\rho_2}\frac{h_1 h_2}{h_1+h_2}}\,,
\label{eq:cphi_2layer}
\end{equation}
as can be seen in Fig.~\ref{fig:Fv_Ekman} where the value of $c_\phi^m$ is indicated by the circle.
A typical evolution of the drag versus speed obtained by Ekman is shown in this figure. The experimental points are crosses and are compared to linear theory of viscous drag in steady state (continuous lines). The local maximum of the drag for the stratified case is reached for a speed slightly smaller than $c_\phi^m$. At high speeds compared to $c_\phi^m$, the drag is similar to the quadratic law for homogeneous fluids, dominated by viscous drag.

This behavior has been reproduced since then, and similar results were obtained by \cite{bib:Milohetal93} or \cite{bib:Vosperetal99} for instance.

% REAR Waves %
ii) Another contribution of Ekman concerns the description of the interfacial waves generated at the rear of the boat. Two types of waves, transverse and divergent, could be observed.
It must be noted that only interfacial waves can be easily observed since the surface waves associated are of very weak amplitude, their amplitude being related to the interfacial waves ones with a ratio equals to ($\rho_{a}-\rho_1)/(\rho_2-\rho_1)\simeq1/500$ (where $\rho_a$ is the density of air). The free surface remains still at the laboratory scale.

These waves are generated by a depression that develops at the rear of the boat while moving. The transverse waves can reach large amplitudes up to breaking. But they tend to disappear when the speed of the boat is greater than $c_\phi^m$, where only divergent waves remain.

Visualizations of these waves from numerical simulations can also be observed in~\citep{bib:Milohetal93,bib:Yeung99}. Their properties are set by the Froude number only, which compares the mean speed of the moving object to $c_\phi^m$. This will help us in our description later on.

% BOW Wave %
iii) Furthermore, a solitary wave at the bow of the boat can also be observed. This structure, a spatially localized bump is reminiscent of solutions of the Korteweg-DeVries (KdV) equation: it can evolve freely, conserving its shape when the boat is stopped. Otherwise, it remains trapped to the boat.

% HYSTERESIS behavior %
iv) From these observations, Ekman gave an interpretation of the hysteretic behavior the speed of a boat can experience when evolving in dead-water. The analysis can
be explained with the help of Fig.~\ref{fig:Fv_Ekman} showing the drag-speed relation established by Ekman from linear theory. If the force moving the boat increases such that the boat accelerates from rest, the speed of the boat can jump from $6$ to $15$~cm.s$^{-1}$ when the boat overcomes the maximum drag. Similarly, when the force diminishes such that the boat decelerates from a speed larger than $c_\phi^m$, a sudden decrease from $11$ to $4$~cm.s$^{-1}$ occurs. The range of values between $6$ and $11$~cm.s$^{-1}$ is thus an unstable branch inaccessible to the system, where dead-water phenomenon occurs.

However it is important to emphasize that this analysis implies changing the moving force, hence not imposing a constant one.
This observation is different from another remark made by Ekman which relates explicitly to the apparent unsteady behavior of the boat while towed by a constant force~\citep[][p.~67]{bib:Ekman1904}. He observed oscillations of the speed of the boat that could be of large amplitude compared to the mean value.
He noticed that they occur when the boat is evolving at speeds smaller than $c_\phi^m$, while amplitudes and periods of the oscillations depend on the towing force and the properties of the stratification.
We emphasize here that this last property is not included in analytical approaches considering linear waves, such as~\citep{bib:Ekman1904,bib:Milohetal93,bib:Yeung99}, and seems to be an important characteristic of the dead-water phenomenon.

%% Other dead-water experiments %%
\subsection{Other dead-water related works}
\cite{bib:Hughes78} took advantage of the dead-water effect to study the effects of interfacial waves on wind induced surface waves. The study relates the statistical properties of surface waves to the currents induced by internal waves.

\cite{bib:Maas06} investigated if the dead-water effect could also be experienced by swimmers in a thermally stratified pool, offering a plausible explanation for unexplained drownings of experienced swimmers in lakes during the summer season, but found no effect. One can argue that the stratification considered might have been inadequate for swimmers to generate waves and led to mixing of the thermocline mainly. An energetic budget is given in a more detailed and idealized study~\citep[][]{bib:Maasetal09} where the authors also observed some retarding effects on the swimmers.

In a slightly different perspective, \cite{bib:Nicolaou95} demonstrated that an object accelerating in a stratified fluid generates oblique and transverse internal waves, the latter can be decomposed as a sum of baroclinic modes with the lowest mode always present. \cite{bib:Shishkina02} further showed through experiments that in such a dynamical evolution, the baroclinic modes generated propagate independently of each other, although nonlinear effects must become important when the amplitude of the internal waves is increasing.

\subsection{Steady motion, body moving at constant speed}

Finally, numerous studies were focused on bodies evolving at constant speed {\em within} a stratified fluid. Some results can help the understanding of the dead-water experiments, especially  the internal waves at the rear of the boat and the wave-induced drag on the boat.

In the case of a two-layer fluid, the drag on the boat is maximal when the Froude number, defined as the ratio of the boat speed $U$ to the maximum wave speed given in Eq.~(\ref{eq:cphi_2layer}), is slightly less than 1~\citep{bib:Milohetal93,bib:Motygin97}: this is the subcritical regime. The structure of the internal waves generated and their coupling with surface waves confirm that  the dead-water regime is due to baroclinic waves only~\citep{bib:Yeung99}.

These results obtained in a linear case can be extended when considering weakly nonlinear effects~\citep{bib:Baines1995}. Fully nonlinear calculations are needed when the amplitude of the waves reaches about 0.4 times the depth of the thinner layer~\cite{bib:Grueetal97,bib:Grueetal99}.

In the case of a linearly stratified fluid, the drag is again maximal for slightly subcritical values of the Froude number~\citep{bib:Greenslade00}. Nevertheless, the internal waves emitted can be very different depending on the regime considered for the Froude number~\citep{bib:Chomazetal93}, the location of the object being at the surface~\citep{bib:RottmanetalAPS2004} or fully immersed~\citep{bib:Hopfingeretal91, bib:Meunieretal06}.
It is interesting to emphasize that in experiments done at constant speed, difficulties are often encountered to reach a steady state, as noticed by \cite{bib:Vosperetal99} for instance.

Nevertheless, the dead-water phenomenon does not correspond to a constant speed evolution. By imposing a constant force to move the boat, or using a motor at constant power, the speed of the boat is free to evolve.
The dynamical study of the problem is thus much richer than in the steady state case.

%%%%%%%%%%%%%%%%%%%%%%%%%%%%%%%%%%%%%%%%%%%%%%%%
%%%%%%%%%%% EXPERIMENTAL SETUP %%%%%%%%%%%%%%%%%
\section{Experimental setup}
\label{experimentalsetup}

\begin{figure}[!htb]
\begin{center}
\begin{picture}(10,2.5)
\put(0.25,-0.5){\includegraphics[width=8\unitlength]{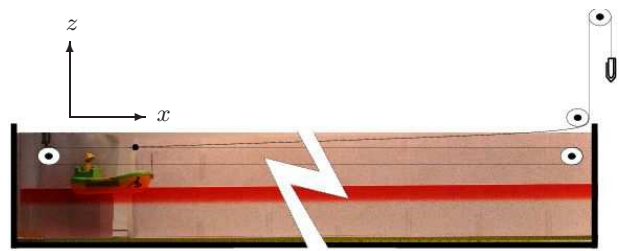}}
\put(1,2.5){$z$}
\put(2.2,1.3){$x$}
\put(1.05,1.35){\vector(1,0){1}}
\put(1.05,1.35){\vector(0,1){1}}
\end{picture}
\end{center}
\caption{Experimental setup of a boat dragged by a falling weight. The vertical force is converted horizontally through pulleys and an horizontal belt of fixed tension.}\label{fig:eauxmortes_montage}
\end{figure}

Nowadays technologies give us the opportunity to gain more insight into the interactions between the interfacial waves and the boat, and improve our knowledge of the dead-water phenomenon.
The experimental setup is described in Figure~\ref{fig:eauxmortes_montage}.

We drag a plastic Playmobil\,\copyright\, boat of width $10$~cm (and with a fisherman and fishes to modulate its weight) with a falling weight in a $3$-m long plexiglass tank of width $10.5$~cm filled with a stratified fluid.
A belt with fixed tension is used to move the boat with a constant horizontal force. The tension of the belt is set to a constant value throughout all the experiments.
The falling weight of mass $m$ fixed to the belt sets the boat into motion, such that the constant force used is gravity $mg$. One must notice that the weights used are paper clips of few milligrams, since a very small force is required to propel the boat within the interesting regime.
A magnet fixed to the boat is used to release it in a systematic way thanks to an electro-magnet outside the tank.

\subsection{Stratification}
Different types of stratification have been used. The density profiles presented here are obtained using a conductivity and temperature probe (CT-probe) from PME\,\copyright.

\begin{figure}[!htb]
\begin{center}
\begin{picture}(10,4.5)
\put(0.5,0){\includegraphics[width=8\unitlength]{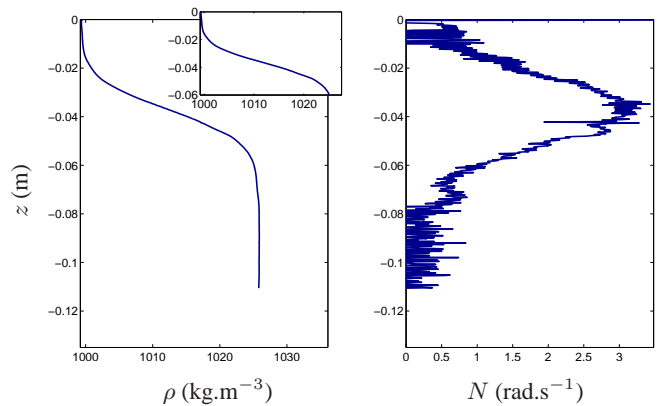}}
\put(0,1.9){\rotatebox{90}{$z$ (m)}}
\put(2,-0.5){$\rho$ (kg.m$^{-3}$)}
\put(6,-0.5){$N$ (rad.s$^{-1}$)}
\end{picture}
\end{center}
\caption{Density profile $\rho$ and Brunt-V\"{a}is\"{a}l\"{a} frequency $N$ for a two-layer fluid after a series of experiments. The density jump is approximately $4$~cm wide as observed in the zoomed window close to the free surface.}\label{fig:bicouche_profile}
\end{figure}

\begin{figure}[!htb]
\begin{center}
\begin{picture}(10,4)
\put(0.5,0){\includegraphics[width=8\unitlength]{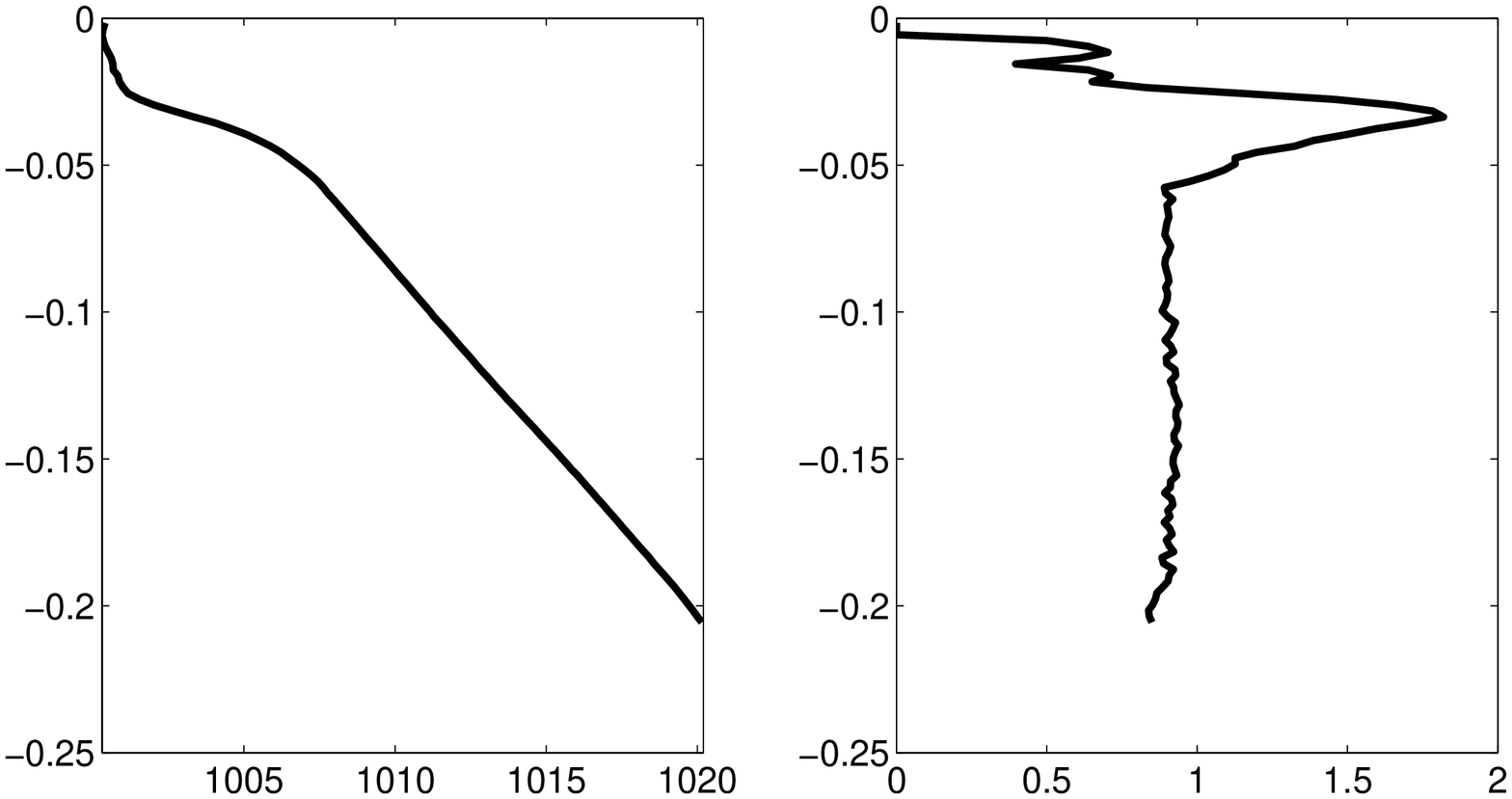}}
\put(0,1.9){\rotatebox{90}{$z$ (m)}}
\put(2,-0.5){$\rho$ (kg.m$^{-3}$)}
\put(6,-0.5){$N$ (rad.s$^{-1}$)}
\end{picture}
\end{center}
\caption{Density profile $\rho$ and Brunt-V\"{a}is\"{a}l\"{a} frequency $N$ obtained experimentally in the case of a linear stratification with a pycnocline.}\label{fig:eauxmortes_linear_profil}
\end{figure}

1) The two-layer fluid is composed of a layer of fresh water (density $\rho_1$) colored with red food dye resting above a transparent layer of salt water (density $\rho_2>\rho_1$).
By siphoning the interface, the density jump extension can be reduced to a few millimeters but through successive experiments, diffusion and mixing make it widen with time, up to a few centimeters as can be seen in Fig.~\ref{fig:bicouche_profile}.

ii) The three-layer fluid is obtained similarly, by adding another layer of salt fluid (density $\rho_3>\rho_2$) colored with green food dye from the bottom.

iii) Finally, a continuously stratified fluid composed of an homogeneous layer above a linearly stratified part can be obtained from adding a fresh layer after filling the tank with the classic ``two-bucket" method~\cite[]{bib:Hill02}. An example of such stratification with pycnocline is shown in Fig.~\ref{fig:eauxmortes_linear_profil}, along with the Brunt-V\"{a}is\"{a}l\"{a} frequency associated with it.

% Measurement techniques
% Comments on checking the interface position with time through schlieren.
\subsection{Techniques of image analysis}

\begin{figure}[!htb]
\begin{center}
\begin{picture}(10,5)
\put(0.4,2.6){\includegraphics[width=8.01\unitlength]{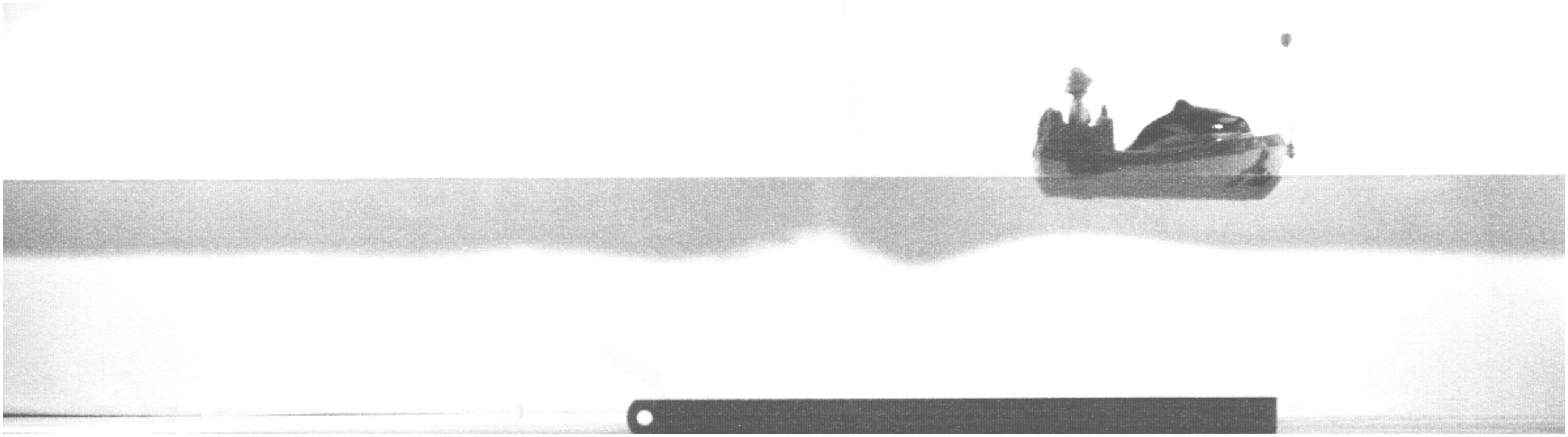}}
\put(0.5,4.4){(a)}
\put(0.4,0){\includegraphics[width=8\unitlength]{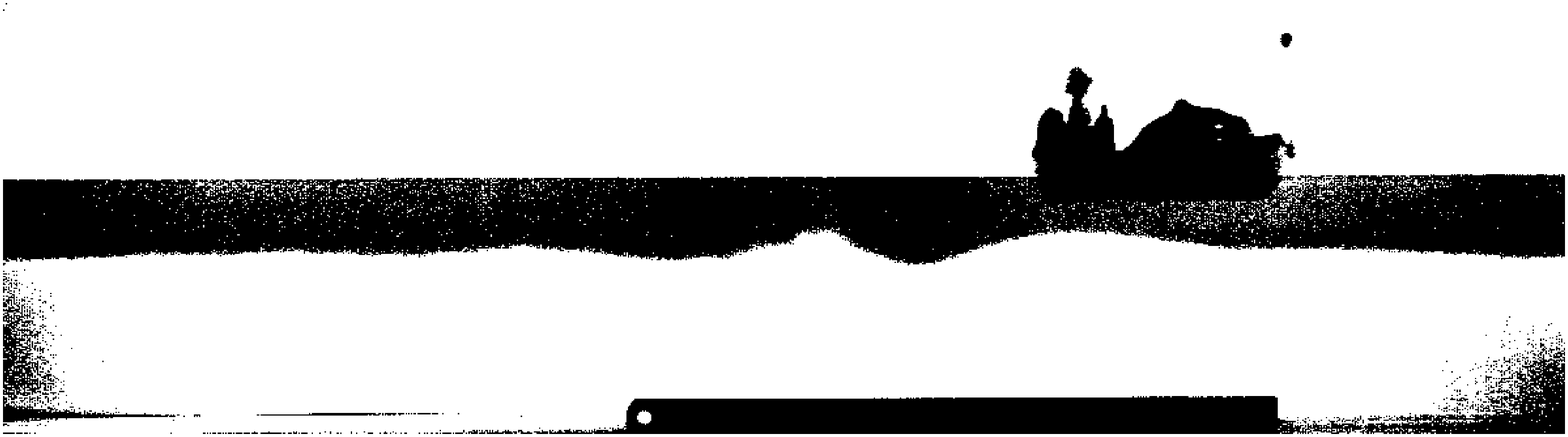}}
\put(0.5,1.8){(b)}
\put(0,0.75){$z$}
\put(1.2,-0.45){$x$}
\put(0.05,-0.4){\vector(1,0){1}}
\put(0.05,-0.4){\vector(0,1){1}}
\end{picture}
\end{center}
\caption{(a) Gray scale image during an experiment, converted into a (b) two-level black and white image with technique 1.}\label{fig:eauxmortes_technique1}
\end{figure}

We record the dynamics of the system (waves + boat) using a black and white camera. Depending on the stratification considered, two different techniques are used to extract the dynamics.
% 2-3 layer case
For the two and three-layer cases, the different layers are identified using food dye and the position of the interfaces $\eta(x,t)$ are extracted from highly contrasted images, as can be seen in Fig.~\ref{fig:eauxmortes_technique1}. The boat position with time $x(t)$ and the free surface evolution are also obtained from these images.
Several hypothesis must be highlighted when dealing with information obtained with this technique, that we called technique~$1$:
\begin{itemize}
\item we consider the interface between two layers as infinitely thin,
\item we neglect diffusion of salt and dye, which is a slow phenomenon compared to the experiments (typically a day versus a few minutes),
\item we neglect the small scale evolutions of the interface, more specifically mixing events.
\end{itemize}

The second technique used, technique~$2$, is based on synthetic schlieren~\citep[][]{bib:Dalzieletal00, bib:Dalzieletal07} and can be realized simultaneously with technique~$1$.
By computing correlations between images in the stratified case and a reference image with homogeneous water,  one can measure the complete density gradient and not only the one associated with the internal waves field.
In the domain considered, one can thus access $\partial_x\rho(x,z,t)$ and $\partial_z\rho(x,z,t)$. In the absence of any motion, the second quantities have been integrated to measure the density profile and it was compared successfully to the one obtained with the CT-probe.
Allowing to quantify the large oscillations of strong density gradient, this technique has been especially considered in the case of a continuous stratification with a pycnocline.

\begin{figure}[!ht]
\begin{center}
\begin{picture}(10,5.2)
\put(0.52,-0.1){\includegraphics[width=5.975\unitlength]{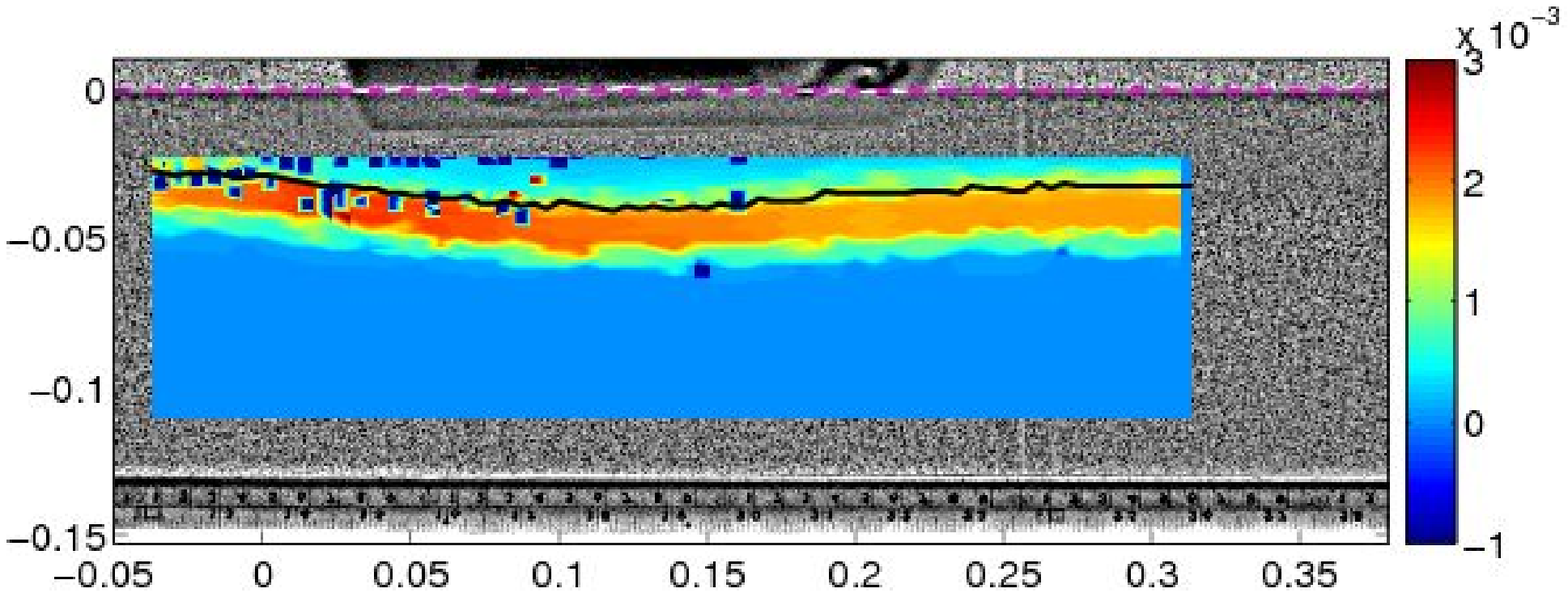}}
\put(0,0.65){\rotatebox{90}{$z$ (m)}}
\put(2.75,-0.5){$x$ (m)}
\put(6.5,0.9){$\partial_z\rho$ (g~cm$^{-4}$)}
\put(1,2.2){(b)}
\put(0.5,3){\includegraphics[width=8\unitlength]{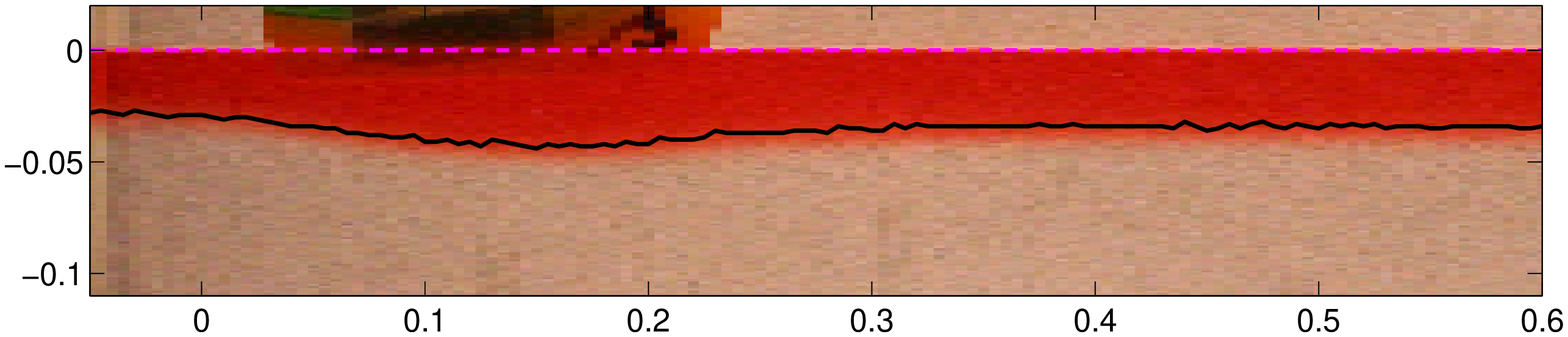}}
%\put(0.5,3){\includegraphics[width=8\unitlength]{images_eauxmortes/2layer_exp2_6tr_canon_im7_old}}
\put(0,3.5){\rotatebox{90}{$z$ (m)}}
\put(4,2.6){$x$ (m)}
\put(1,4.9){(a)}
\end{picture}
\end{center}
\caption{Technique 2 for a two-layer fluid experiment: (a) color image (taken with camera) and (b) gray scale image over which is superposed the vertical gradient of density $\partial_z\rho$ in g~cm$^{-4}$, obtained using synthetic schlieren with a reference image with homogeneous water. The black line corresponds to the interface position extracted from image (a) and added to image~(b) while the pink dashed line is the free surface.}\label{fig:eauxmortes_technique2}
\end{figure}

We have checked that technique~$1$ gives a good indication of the position of the interface by recording synchronously a two-layer case with both techniques.
As can be seen in Fig.~\ref{fig:eauxmortes_technique2} where the image~(a) is recorded with technique~$1$ and image~(b) with technique~$2$, the position of the interface (black line) extracted from the image~(a) follows the main evolution of the density jump in (b) where the vertical density gradient is the strongest.

Although technique~$2$ can be used for all experiments, we will use technique~$1$ to analyze the two and three-layer cases since its computational cost is much less.

%%%%%%%%%%%%%%%%%%%%%%%%%%%%%%%%%%%%%%%%%%%%%%%%%
%%%%%%%%%%%%%% 2 LAYER CASE %%%%%%%%%%%%%%%%%%%%%
\section{Revisiting Ekman's work}
\label{twolayer}

Our first experimental investigation is dedicated to the two-layer case, in order to reproduce the results obtained by Ekman.
The main parameters are summarized in Table~\ref{tab:params_eauxmortes}.
\begin{table*}[!htb]
%\begin{center}
%\hspace{-1cm}
\begin{tabular}{llccl}
\multicolumn{2}{l}{Parameters} & symbols & values & units \\
\hline\\
Tank & dimensions & $L\,\times\,H\,\times\,W$ & $300\,\times\,50\,\times\,10.5$ & cm$^{3}$ \\
\\
\multirow{2}{*}{Traction} & belt tension & $T$ & $\sim 0.35$ & N \\
		& force & $F_t$ & $0.011 - 0.035$ & N\\ 
\\
\multirow{3}{*}{Boat} & dimensions & $L_b\,\times\,h_b\,\times\,w_b$ & $20.0\,\times\,5.0\,\times\,10.0$ & cm$^{3}$ \\
		& immersed section & $S_b$ & $12.0 - 24.0$ & cm$^{2}$ \\
		& mass & $M$ & $171 - 343$ & g \\
\\
\multirow{2}{*}{Fluid 1} & density & $\rho_1$ & $0.999 - 1.005$ & g~cm$^{-3}$ \\
		& depth & $h_1$ & $2.0 - 5.0$ & cm\\ 
\\
\multirow{2}{*}{Fluid 2} & density & $\rho_2$ & $1.010 - 1.030$ & g~cm$^{-3}$ \\
		& depth & $h_2$ & $5.0 - 15.0$ & cm\\ 
\\
 & mean density & $\bar{\rho}=\frac{\rho_2+\rho_1}{2}$ & $\sim 1.01$ & g~cm$^{-3}$\\
 & density jump & $\Delta\rho=\frac{\rho_2-\rho_1}{\bar{\rho}}$ & $0.01 - 0.1$ & \\
 & maximum phase speed & $c_{\phi}^m=\sqrt{\Delta\rho g\frac{h_1 h_2}{h_1+h_2}}$ & $0.03-0.2$ & m~s$^{-1}$\\
 & Froude number & $Fr=\frac{U}{c_{\phi}^m}$ & $0.2-2$ & \\
 & Reynolds number & $Re=\frac{U h_1}{\nu}$ & $400-10000$ & \\
\end{tabular}
%\end{center}
\caption{Experimental parameters used for the experiments with a two-layer fluid.}\label{tab:params_eauxmortes}
\end{table*}

\subsection{Dynamics of the boat}

As can be seen in Table~\ref{tab:params_eauxmortes}, we have used several sets of parameters in order to verify that the dead-water phenomenon is indeed a function of the relative density jump $\Delta\rho=({\rho_2-\rho_1})/{\bar{\rho}}$, the fresh layer depth $h_1$ compared to the salted one $h_2$ along with the waterline $h_b$ on the boat (changing $h_b$ implies a different immersed cross-section and thus a different viscous drag). In the following, we will refer to the immersed cross-section $S_b$ for varying geometry of the boat, and which corresponds to $w_b\times h_b$.

\begin{figure}[htb]
\begin{center}
\begin{picture}(10,5.2)
\put(0.5,2.75){\includegraphics[width=8\unitlength]{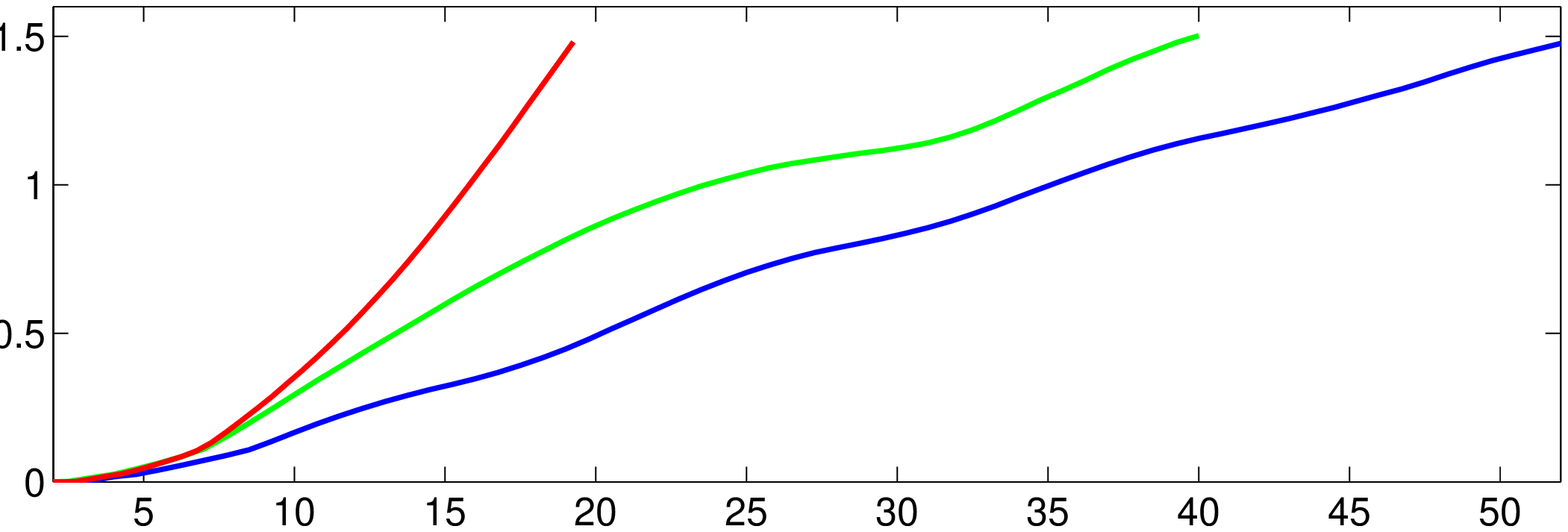}}
\put(0,3.85){\rotatebox{90}{$x(t)$}}
\put(1.2,4.85){(a)}
\put(0.5,0){\includegraphics[width=8\unitlength]{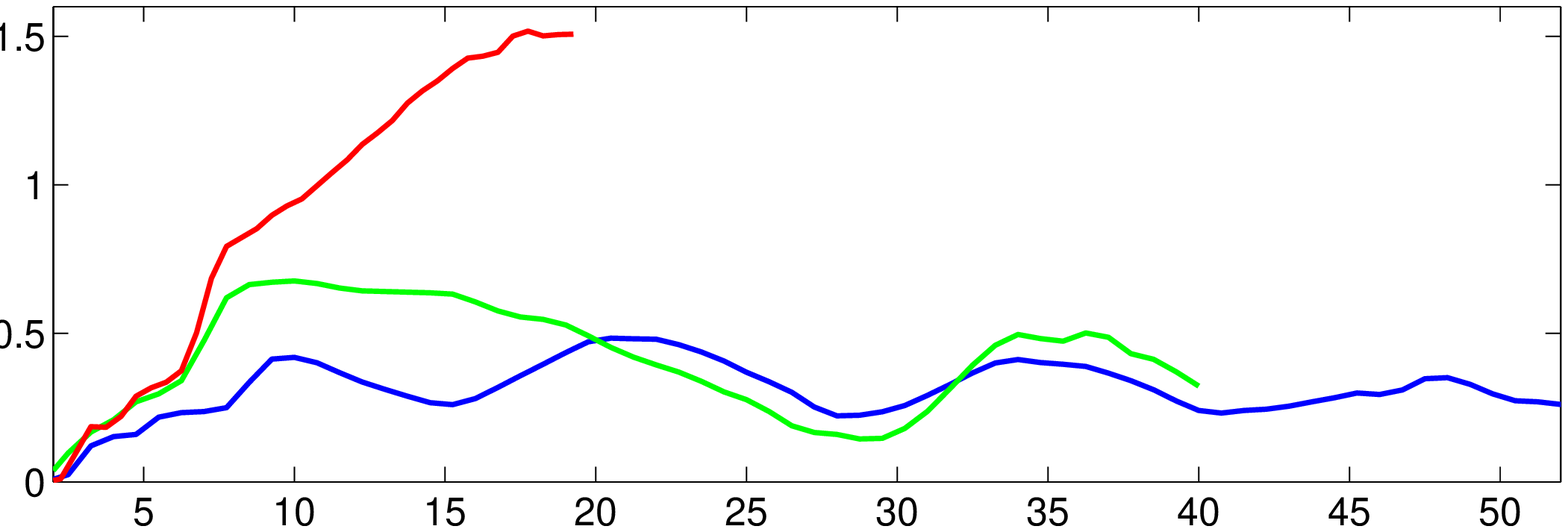}}
\put(0,1.1){\rotatebox{90}{$v(t)/c_{\phi}^m$}}
\put(4.25,-0.5){$t$ (s)}
\put(1.2,2.1){(b)}
\end{picture}
\end{center}
\caption{(a) Position of the boat $x(t)$ (in m) and (b) its speed compared to the maximum phase speed $v(t)/c_{\phi}^m$ for experiments revealing the dead-water phenomenon (blue and green) and for an experiment without oscillation (red).}\label{fig:xdxdt_deadwater}
\end{figure}

For instance, the dead-water phenomenon is easy to reproduce with the following parameters: $h_1=5$~cm, $h_2=14$~cm, $h_b\simeq 2.4$~cm,  $\Delta\rho=0.0244$ and the falling weight corresponding to a force $F_t=16.3$~mN (also studied in Sec.~\ref{sec:2layerXTdiagram}, Fig.~\ref{fig:eauxmortes_oscillations}).
The time evolution of the boat $x(t)$ and its corresponding oscillating speed $v(t)=dx/dt$ normalized by $c_{\phi}^m$, associated with this experiment, are shown in blue in Fig.~\ref{fig:xdxdt_deadwater}.
A similar case but with a force $F_t=18.8$~mN (also studied in Sec.~\ref{sec:2layerXTdiagram}, Fig.~\ref{fig:Fr_o1_reflab}) is shown in green in Fig.~\ref{fig:xdxdt_deadwater}.

Ekman already commented on the fact that the speed can have large fluctuations $|v_{max}-v_{min}|\sim 0.027$ m.s$^{-1}$ compared to its mean value $<v>\simeq 0.031$ m~s$^{-1}$ (blue case), referring to the capricious nature of the phenomenon.
% values for exp 9 (dv~0.027, vm~ 0.031)
% values for exp 10 (dv~0.05, vm~ 0.042)
Oscillations of the order of $85\%$ of the mean value make it difficult to interpret the dead-water regime as stationary.
The force imposed to move the boat being constant, such a temporal evolution implies a time-evolving drag due to the interfacial waves, and that is due to generation and breaking of waves at the rear of the boat.
This will be confirmed later on when we present the spatio-temporal diagram associated with the interface position.

We also show in red in Fig.~\ref{fig:xdxdt_deadwater} the time evolution of $x(t)$ and $v(t)/c_{\phi}^m$ for the same parameters except $F_t=21$~mN (also studied in Sec.~\ref{sec:2layerXTdiagram}, Fig.~\ref{fig:Fr_o1_reflab}). Although only reached at the end of the recording, the boat tends to a steady evolution at constant speed, $v_\ell\simeq0.14$~m~s$^{-1}$, after a transient state (for $t\leq15$~s).
The comparison between the red and green case also reveals the sensitivity of the system regarding the drag force. The temporal evolutions are initially very similar (for the first five seconds) before behaving differently.
For sufficiently large speeds, the hypothesis of constant velocity of the boat seems reliable and can lead to a good interpretation of the evolution of the boat.

\begin{figure}
\begin{center}
\begin{picture}(10,3.25)
\put(0.5,0){\includegraphics[width=8\unitlength]{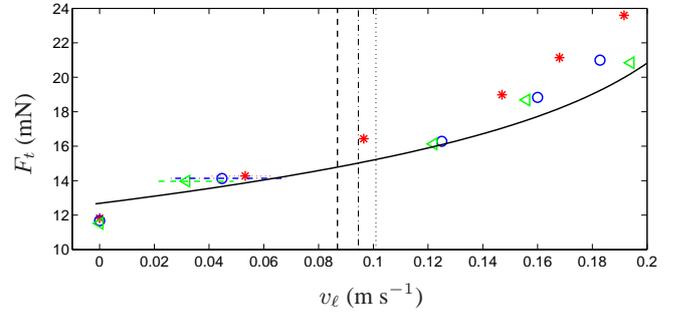}}
\put(4,-0.5){$v_\ell$ (m~s$^{-1}$)}
\put(0,1.2){\rotatebox{90}{$F_t$ (mN)}}
\end{picture}
\end{center}
\caption{Speed-Force diagram for three stratifications with a fresh layer $h_1$ being respectively $(\ast)0.06$~m, $(\circ)0.05$~m and $(\lhd)0.04$~m; with a salt layer $h_2=0.14$~m and $\Delta\rho=0.0247$, for a boat configuration being $S_b=12$~cm$^2$. For the unsteady regime, horizontal lines are drawn corresponding to the range of values of the speed in which it fluctuates. The continuous line corresponds to a homogeneous case of depth $0.15$~m. The vertical dashed lines gives the values of $c_\phi^m$ associated with $h_1$, (dotted line) 0.10~m~s$^{-1}$, (dash-dotted line) 0.095~m~s$^{-1}$ and (dashed line) 0.087~m~s$^{-1}$.}\label{fig:Fv_3h1}
\end{figure}

\begin{figure}
\begin{center}
\begin{picture}(10,3.25)
\put(0.5,0){\includegraphics[width=8\unitlength]{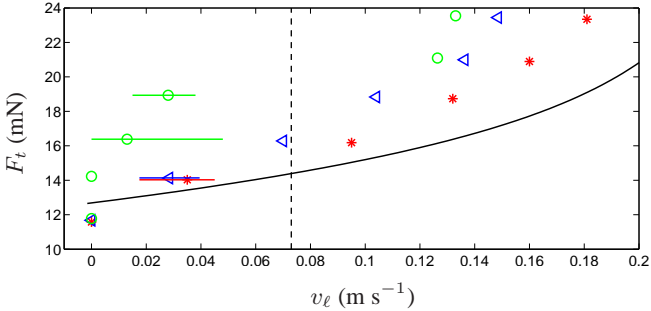}}
\put(4,-0.5){$v_\ell$ (m~s$^{-1}$)}
\put(0,1.2){\rotatebox{90}{$F_t$ (mN)}}
\end{picture}
\end{center}
\caption{Speed-Force diagram for three boat configurations, $(\ast)-S_b=12$,  $(\lhd)-S_b=18$,  and $(\circ)-S_b=24$~cm$^2$ in a stratification with $h_1=0.06$~m, $h_2=0.10$~m and $\Delta\rho=0.0145$. For the unsteady regime, horizontal lines are drawn corresponding to the range of values of the speed in which it fluctuates. The continuous line corresponds to a homogeneous case of depth $0.15$~m with $S_b=18$~cm$^2$. The vertical dashed line gives the value of $c_\phi^m$}\label{fig:Fv_3boats}
\end{figure}

Analogously to Ekman's results presented in  Fig.~\ref{fig:Fv_Ekman}, and in order to compare our results with his, we summarize in Figs.~\ref{fig:Fv_3h1} and.~\ref{fig:Fv_3boats} several time evolutions for the different stratifications and boat configurations considered in a speed-force diagram.
However, two comments must be made with respect to Ekman's presentation of his experimental results. When the boat is in an oscillating regime, Ekman takes the mean value as its speed and considers that the drag equals the moving force. This simplified representation of the data leads to cleaner diagrams although some strong hypotheses are hidden.

In our case, we consider the moving force $F_t$ instead of the drag force since both forces are equal only when the boat evolution is steady.
We represent the speed by its mean value (symbols) and the range of values (lines) in which it fluctuates.% when no steady state is reached.

Figure~\ref{fig:Fv_3h1} combines the speed-force diagram associated with three different stratifications and one boat configuration ($S_b=12$~cm$^2$), whereas Fig.~\ref{fig:Fv_3boats} gathers three boat configurations and one stratification.

These two diagrams are in agreement with the established results by Ekman that the wave-induced drag increases along with the fresh water depth $h_1$ when a steady state exists (Fig.~\ref{fig:Fv_3h1}), and depends also on the boat geometry (Fig.~\ref{fig:Fv_3boats}).
As the immersed section of the boat is larger, the drag gets stronger. The best coupling occurs for $h_b\simeq h_1/2$ where the wave induced drag is maximum and the boat is not deep enough to generate mixing at the interface.
Furthermore, both diagrams show that when the speed of the boat is oscillating, it never reaches values larger than $c_\phi^m$.
It is important to notice that the dynamical regime where the velocity of the boat is closed to $c_\phi^m$ is difficult to scan since the tiniest change of the moving force leads to a completely different evolution of the boat. One can thus argue there exists a range of ``unattainable" values for the velocity of the boat, as discussed in Sec.~\ref{sec:EkmanPhD}, \S~iv. However, since the moving force is kept constant in our case, this aspect is different from Ekman's presentation of the phenomenon.
Similar comparisons have been made for the influence of the density jump $\Delta\rho$ and the salt layer depth $h_2$ but are not shown here for the sake of concision.

Finally, several general comments are in order
\begin{itemize}
\item the wave-induced drag is always stronger than the one in homogeneous water (viscous drag only),
\item  a regime of low velocities exists with a non constant drag and large fluctuations in the boat speed (no steady state can be reached),
\item this regime is always associated with velocities smaller than $c_\phi^m$.
\end{itemize}
From these remarks, we can discriminate the two regimes described previously according to the Froude number $Fr$ which is defined as the ratio of the speed of the boat (that can be its mean value $<v>$ or its limit $v_\ell$) to the maximum phase speed of the interfacial waves.
It is important to realize here that the Froude number is intrinsically a time evolving parameter, but we will use characteristic values to identify the regime considered.

We now focus on the waves according to this classification.

\subsection{Interfacial waves}
\label{sec:2layerXTdiagram}

We present here observations of the interfacial position $\eta(x,t)$ generated by the boat in two different frames, the laboratory reference frame and the one moving with the boat. The latter is not Galilean since the boat does not evolve at constant speed, but offers a better understanding of the coupled dynamics of the waves with the boat.

Visualizations of the interfacial positions are represented in spatio-temporal diagrams where elevations of the interface are in red and depressions in blue (see Figs.~\ref{fig:Fr_O1_reflab} and \ref{fig:Fr_o1_reflab} for instance).
The bow and stern of the boat are superimposed on these diagrams (as black lines).

Depending on the Froude number being larger or smaller than $1$, the wave dynamics is very different. We illustrate each case with an example corresponding to a stratification with  $h_1=5.0$~cm, $h_2=14.0$~cm, $\rho_1=0.9980$~g~cm$^{-3}$, $\rho_2=1.0227$~g~cm$^{-3}$, leading to $c_\phi^m=0.094$~m~s$^{-1}$. The boat used corresponds to an immersed section $S_b=24$~cm$^2$.

\subsubsection{In the laboratory frame}
\label{sect:eauxmortes_bicouche_reflabo}

\begin{figure}
\begin{center}
\begin{picture}(10,5.25)
\put(0.5,-0.1){\includegraphics[width=8\unitlength]{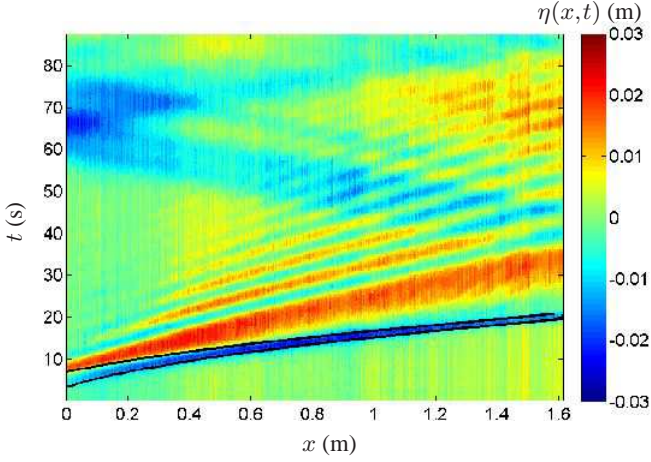}}
\put(3.9,-0.5){$x$ (m)}
\put(0,2.25){\rotatebox{90}{$t$ (s)}}
\put(7,5.2){$\eta(x,t)$ (m)}
\end{picture}
\end{center}
\caption{{\em Ballistic regime} ({\it $Fr>1$}). Spatio-temporal diagram of interfacial displacements $\eta(x,t)$ (in m). The solid black lines are the bow and stern of the boat. Experimental parameters: $h_1=5.0$~cm, $h_2=14.0$~cm, $\rho_1=0.9980$~g~cm$^{-3}$, $\rho_2=1.0227$~g~cm$^{-3}$, $S_b=24$~cm$^2$, $F_t=21$~mN.}
\label{fig:Fr_O1_reflab}
\end{figure}

Let us first consider the case where a steady state is reached, illustrated in Fig.~\ref{fig:Fr_O1_reflab}. With a moving force $F_t=21$~mN, the limit speed reached by the boat is $v_\ell\simeq0.15$~m~s$^{-1}$ which gives $Fr=1.6$.
In this supercritical case, one can observe that the boat escapes from the wave train it generates at start-up.
A depression (the blue region) remains attached below the boat and propagates with it at the speed of the boat.
For times larger than $40$~s, the boat meets with the end of the tank and stops (around $x=2.5$~m). The depression below it is thus expelled and propagates in the other direction after reflecting on the side of the tank. This case is in the ballistic regime.

\begin{figure}
\begin{center}
\begin{picture}(10,5.25)
\put(0.5,-0.1){\includegraphics[width=8\unitlength]{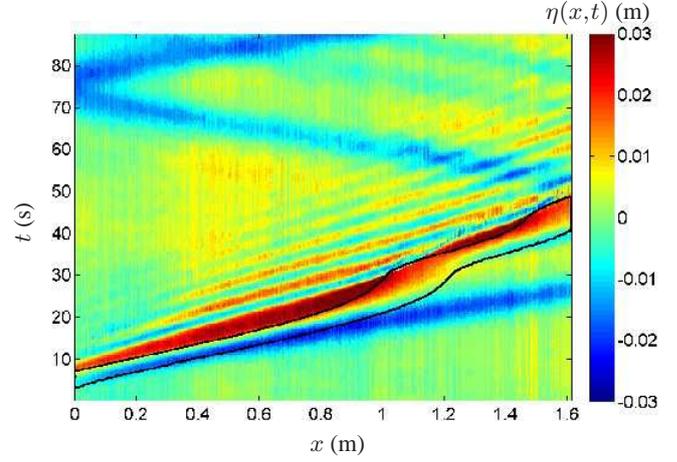}}
\put(3.9,-0.5){$x$ (m)}
\put(0,2.25){\rotatebox{90}{$t$ (s)}}
\put(7,5.2){$\eta(x,t)$ (m)}
\end{picture}
\end{center}
\caption{{\em Oscillating regime} ({\it $Fr< 1$}). Spatio-temporal diagram of interfacial displacements $\eta(x,t)$ (in m). The solid black lines are the bow and stern of the boat. Experimental parameters: $h_1=5.0$~cm, $h_2=14.0$~cm, $\rho_1=0.9980$~g~cm$^{-3}$, $\rho_2=1.0227$~g~cm$^{-3}$, $S_b=24$~cm$^2$, $F_t=18.8$~mN.}
\label{fig:Fr_o1_reflab}
\end{figure}

The oscillating regime, more atypical, with oscillations of the boat is illustrated in Fig.~\ref{fig:Fr_o1_reflab} and corresponds to $F_t=18.8$~mN. The mean speed of the boat is $<v>\simeq0.035$~m~s$^{-1}$, which gives a Froude number of $0.4$.
As the boat evolves, the amplitude of the interfacial waves grows up to a maximum value where the waves catch up with the boat and break on its hull, around $t=20$~s. This corresponds to the moment where the drag force is maximal since the boat is almost standing still. The vessel can speed up again after the first wave breaks and the same dynamics is reproduced.
As noted earlier, a depression located below the boat evolves with the boat when it is accelerating. It evolves freely after the boat is stopped ($t>20$~s) and propagates to the right first and to the left for $t>50$~s after reflection at the side of the tank. This depression even bounces a second time on the left side of the tank and its shape remains almost unchanged ($t>75$~s). Although not presented here, we have verified that this solitary wave propagating upstream is well described by a Korteweg-de Vries model since its amplitude is always smaller than $0.4~h_1$~\citep{bib:Grueetal99}. When the boat starts again, the process is repeated and a new depression is generated.

\subsubsection{In the frame of the boat}

More information can be obtained when following the dynamics in the frame associated with the boat. We are more specifically interested in what sets the amplitude and frequency of the oscillations of the boat when $Fr<1$.

\begin{figure}
\begin{center}
\begin{picture}(10,3.5)
\put(0.5,0){\includegraphics[width=8\unitlength]{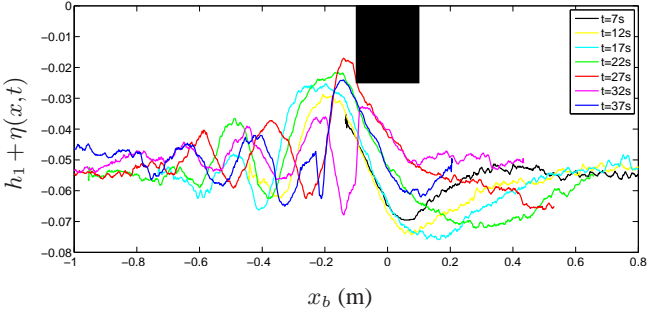}}
\put(4,-0.5){$x_b$ (m)}
\put(0,1){\rotatebox{90}{$h_1+\eta(x,t)$}}
\end{picture}
\end{center}
\caption{{\it $Fr<1$}. Interface evolution $\eta(x,t)$ (in m) from its rest position $h_1$ for several times during the slow down of the boat, in the frame of the boat which is represented by the black rectangular box. Same parameters as in Fig.~\ref{fig:Fr_o1_reflab}.}\label{fig:eauxmortes_raidissement}
\end{figure}

In Fig.~\ref{fig:eauxmortes_raidissement}, we have superposed the interfacial displacements in the frame of the boat (represented as a black rectangle) for different times extracted from Fig.~\ref{fig:Fr_o1_reflab}. We can observe that the waves get closer to the stern of the boat as their amplitude grows. This is mainly due to the decrease of speed of the boat since we can see in Fig.~\ref{fig:Fr_o1_reflab} that the wave crests evolve with a constant speed.
The nonlinear nature of the interfacial waves is also clearly visible as the wave fronts steepen, the limit of growth being set here by the time when the waves hit the hull ($27<t<32$ s).
These downstream features are fully nonlinear (amplitudes larger than $0.6~h_1$) and have similar aspects to observations made in the transcritical regime by \citep{bib:MelvilleHelfrich87} and reproduced numerically by \citep{bib:Grueetal97}, although corresponding here to waves in a non-galilean frame. It is moreover surprising since as observed on the green curve in Fig.~\ref{fig:xdxdt_deadwater}, the Froude number does not exceed 0.7. These observations add to the comments made previously concerning the inadequacy of models based on steady object in stratified fluid.
Here again we see upstream the solitary depression below the boat escaping at the bow as a single oscillation.

For times $t\geq32$~s, one can observe that the decapitated first crest disappears below the boat, the next wave depression takes position below the boat and the initially second elevation of the packet becomes the closest to the stern.
Hence, during an oscillation of the boat, the wave train slides of one wavelength compared to the boat, repeating itself.
No change in the dominant wavelength of the wave train is visible.

We can now estimate the frequency of the oscillations of the boat to be
\begin{equation}
f_b=\frac{c_g}{\lambda}\label{eq:freq_oscillations}
\end{equation}
with $c_g$ the group velocity of the wave train at the rear of the boat and $\lambda$ its dominant wavelength, corresponding approximately to the distance between the first two wave crests.

\begin{figure}
\begin{center}
\begin{picture}(8,6.75)
\put(0.5,3.9){\includegraphics[width=7.5\unitlength]{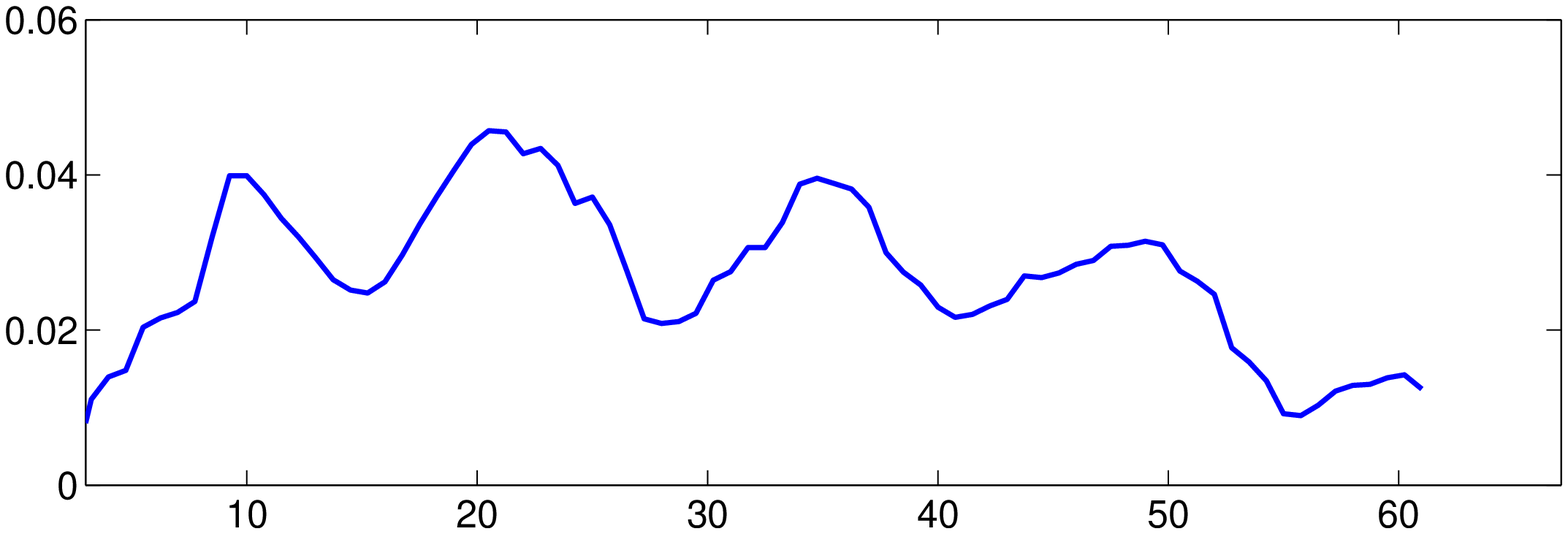}}
\put(4.1,6.5){(a)}
\put(0,4.75){\rotatebox{90}{$v(t)$}}
\put(4,3.6){$t$ (s)}
\put(0.4,-0.25){\includegraphics[width=8\unitlength]{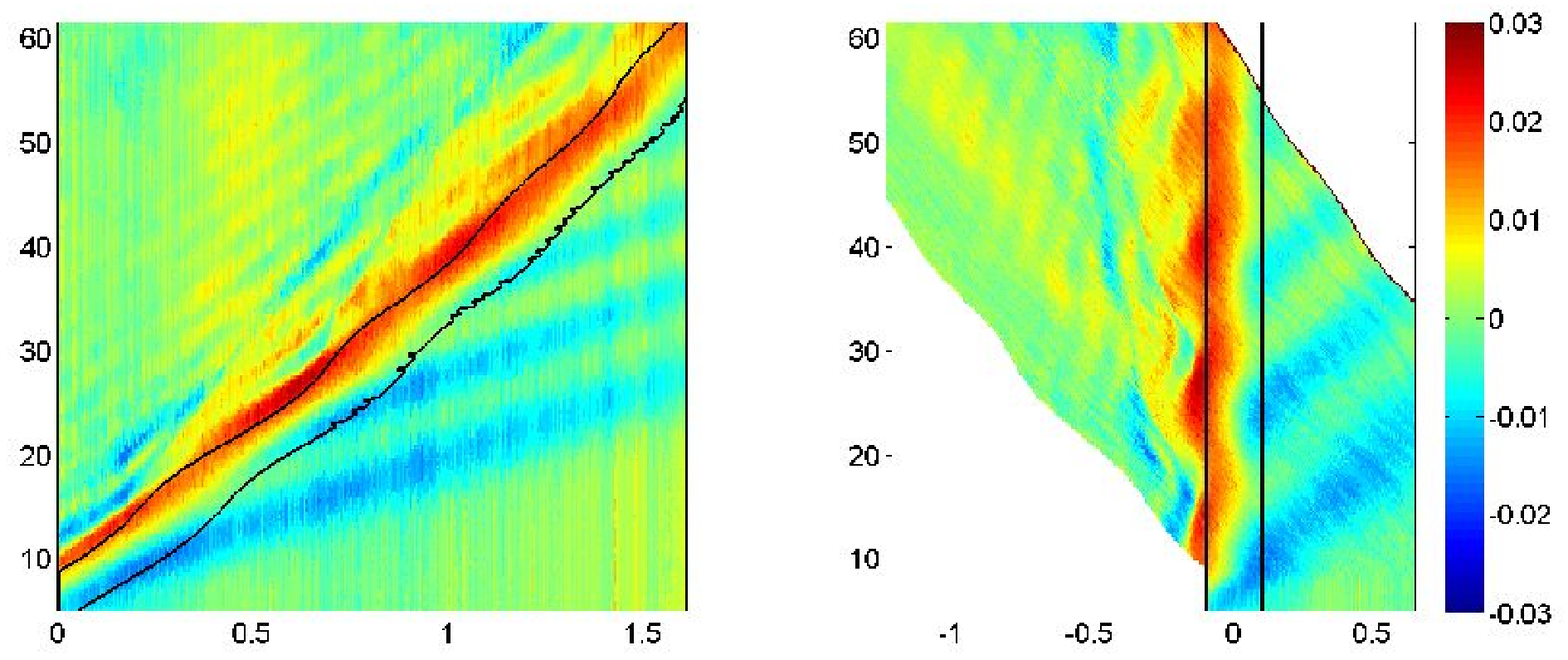}}
\put(2,-0.55){$x$ (m)}
\put(6.2,-0.55){$x_b$ (m)}
\put(2.2,3.1){(b)}
\put(6.3,3.1){(c)}
\put(0,1.2){\rotatebox{90}{$t$ (s)}}
\put(7.2,3.25){$\eta(x,t)$ (m)}
\end{picture}
\end{center}
\caption{{\em Oscillating regime} ({\it $Fr<1$}). Panel (a) presents the speed of the boat versus time, while panels (b) and (c) shows the spatio-temporal diagrams of the interfacial displacements  in the laboratory frame and in the one of the boat respectively. The black lines represent the bow and stern. Experimental parameters: $h_1=5.0$~cm, $h_2=14.0$~cm, $\rho_1=0.9980$~g~cm$^{-3}$, $\rho_2=1.0157$~g~cm$^{-3}$, $S_b=24$~cm$^2$, $F_t=16.3$~mN.}\label{fig:eauxmortes_oscillations}
\end{figure}

The experiment presented in Fig.~\ref{fig:Fr_o1_reflab} allows an observation of the steepening process but does not present many oscillations. In order to confirm the validity of formula~(\ref{eq:freq_oscillations}), we consider another experiment with $F_t=16.3$~mN in Fig.~\ref{fig:eauxmortes_oscillations}.
From Fig.~\ref{fig:eauxmortes_oscillations}~(a), one can estimate the frequency of the oscillations of the boat to be $f_b\sim0.07-0.08$~Hz.
The group speed and wavelength of the wave train obtained from the diagram in Fig.~\ref{fig:eauxmortes_oscillations}~(b), corresponds to
 $c_g\sim0.035-0.040$~m~s$^{-1}$ and $\lambda\sim0.5$~m.
Indeed, we obtain $c_g/\lambda\sim0.07-0.08$~Hz showing very good agreement with $f_b$.

After revisiting experiments similar to the ones performed by Ekman, we have considered new situations.
Indeed, dead-water phenomenon can be studied in more complex (and realistic) stratifications, when the fluid has more than two layers of different densities.
We now turn to new experiments with a three-layer fluid or with a continuously stratified fluid with a pycnocline, where the complex dynamics has also been observed.

%%%%%%%%%%%%%%%%%%%%%%%%%%%%%%%%%%%%%%%%%%%%%%%%%
%%%%%%%%%%%%%% 3 LAYER CASE %%%%%%%%%%%%%%%%%%%%%
\section{Experiments with a three-layer fluid}
\label{threelayer}

\begin{figure*}
\begin{center}
\begin{picture}(18,2.5)
\put(0.25,-0.25){\includegraphics[width=15.5\unitlength]{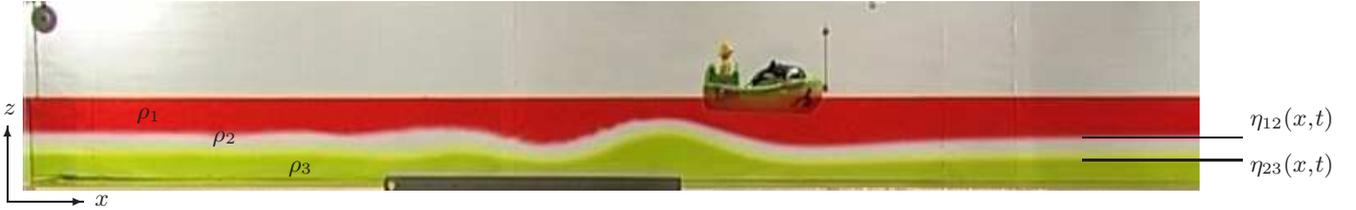}}
\put(0,0.75){$z$}
\put(1.2,-0.45){$x$}
\put(0.05,-0.4){\vector(1,0){1}}
\put(0.05,-0.4){\vector(0,1){1}}
\put(1.75,0.7){$\rho_1$}
\put(2.75,0.4){$\rho_2$}
\put(3.75,0){$\rho_3$}
\put(14.2,0.45){\line(1,0){2.1}}
\put(16.4,0.6){$\eta_{12}(x,t)$}
\put(14.2,0.15){\line(1,0){2.1}}
\put(16.4,0){$\eta_{23}(x,t)$}
\end{picture}
\end{center}
\caption{Picture illustrating the dead-water phenomenon for a three-layer fluid.}\label{fig:eauxmortes_3couches_exemple}
\end{figure*}

We consider a three-layer fluid of depth $h_i$ and density $\rho_i$, $i={1,2,3}$. Two interfaces must be considered now, $\eta_{ij}(x,t)$ corresponding to the interfacial displacements between layers $i$ and $j$. A picture summarizes the setup in Fig.~\ref{fig:eauxmortes_3couches_exemple} and the main parameters are given in Table~\ref{tab:eauxmortes_3couches_params}. Only one given stratification will be discussed here.

\begin{table*}
\begin{tabular}{llccl}
\multicolumn{2}{l}{Parameters} & symbols & values & units \\
\hline\\
\multirow{2}{*}{Fluid 1} & density & $\rho_1$ & $0.9967$ & g~cm$^{-3}$ \\
		& depth & $h_1$ & $5.0$ & cm\\ 
\\
\multirow{2}{*}{Fluid 2} & density & $\rho_2$ & $1.0079$ & g~cm$^{-3}$ \\
		& depth & $h_2$ & $3.0$ & cm\\ 
\\
\multirow{2}{*}{Fluid 3} & density & $\rho_3$ & $1.0201$ & g~cm$^{-3}$ \\
		& depth & $h_3$ & $5.5$ & cm\\ 
\\

\multirow{3}{*}{Interface $\eta_{12}$} & mean density & $\bar{\rho_{12}}=\frac{\rho_2+\rho_1}{2}$ & $1.0023$ & g~cm$^{-3}$\\
 & density jump & $\Delta\rho_{12}=\frac{\rho_2-\rho_1}{\bar{\rho_{12}}}$ & $0.0112$ & \\
 & maximum phase speed & $c_{\phi,12}^m=\sqrt{\Delta\rho_{12} g\frac{h_1 h_2}{h_1+h_2}}$ & $0.0147$ & m~s$^{-1}$\\
 & Froude number & $Fr_{12}=\frac{U}{c_{\phi,12}^m}$ & $4.1-7.5$ & \\
 
\multirow{3}{*}{Interface $\eta_{23}$} & mean density & $\bar{\rho_{23}}=\frac{\rho_3+\rho_2}{2}$ & $1.014$ & g~cm$^{-3}$\\
 & density jump & $\Delta\rho_{23}=\frac{\rho_3-\rho_2}{\bar{\rho_{23}}}$ & $0.012$ & \\
 & maximum phase speed & $c_{\phi,23}^m=\sqrt{\Delta\rho_{23} g\frac{h_2 h_3}{h_2+h_3}}$ & $0.0150$ & m~s$^{-1}$\\
 & Froude number & $Fr_{23}=\frac{U}{c_{\phi,23}^m}$ & $4.0-7.3$ & \\
 
\multirow{2}{*}{Modes $s/a$} & maximum phase speed & $c_{s/a}^m$ & $0.073/0.044$ & m~s$^{-1}$\\
 & Froude number & $Fr_{s/a}=\frac{U}{c_{s/a}^m}$ & $1.5/2.4 - 0.6/0.9$ & \\
\end{tabular}
\caption{Experimental parameters used for the experiments with a three-layer fluid.}\label{tab:eauxmortes_3couches_params}
\end{table*}

We use technique~$1$ again to extract the positions of the two interfaces $\eta_{12}$ et $\eta_{23}$, along with the position of the boat.

\subsection{Analytics}

Although the three-layer case is a simple extension of the well documented two-layer one discussed in previous section, we here reproduce explicit expressions for the phase speed of the two eigenmode solutions of this problem, which can also be find established in the most general form by~\cite{bib:RusasGrue02} or~\cite{bib:Xiao-Gang06} for example.

Perturbations of a three-layer fluid can be described by two types of harmonic interfacial waves, verifying the dispersion relation expressed as the following determinant
\begin{equation}
\left| \begin{array}{cc}
\alpha_{1} & \beta  \\
\beta & \alpha_{2} 
\end{array} \right| = 0\,,\label{eq:det}
\end{equation}
with, for $i={1,2}$
\begin{eqnarray}
\alpha_{i} &=& g(\rho_i-\rho_{i+1})+\frac{\omega^2}{k}\left(\frac{\rho_i}{\tanh (kh_i)} + \frac{\rho_{i+1}}{\tanh (kh_{i+1})} \right),\\
\beta &=& -\frac{\rho_{2}\omega^2}{k\, \sinh (kh_{2})}\,.
\end{eqnarray}
The fourth order polynomial in $\omega$ associated with (\ref{eq:det}) leads to two types of interfacial waves.
Using the notations introduced in Table~\ref{tab:eauxmortes_3couches_params}, and considering the long-wave limit ($\tanh(kh_i)\simeq\sinh(kh_i)\simeq kh_i$) along with the limit of small density jumps ($\Delta\rho_{ij}\ll\rho_i,\rho_j$ and $\rho_i\sim\bar{\rho}\,, \forall i,j$), we can express the two phase velocities associated with each solution
\begin{eqnarray}
c_s^m & = & \left[g\frac{\Delta\rho_{12}h_1(h_2+h_3)+\Delta\rho_{23}h_3(h_1+h_2)+ \tilde{\Delta}}{h_1+h_2+h_3} \right]^{1/2}\,,\label{eq:eauxmortes_mode1}\\
c_a^m & = & \left[g\frac{\Delta\rho_{12}h_1(h_2+h_3)+\Delta\rho_{23}h_3(h_1+h_2)- \tilde{\Delta}}{h_1+h_2+h_3}\right]^{1/2}\,, \label{eq:eauxmortes_mode2}
\end{eqnarray}
with
\begin{eqnarray}
\tilde{\Delta} & = & g\left[\Delta\rho_{12}^2 h_1^2(h_2+h_3)^2+\Delta\rho_{23}^2h_3^2(h_1+h_2)^2 \right. \nonumber\\
& & \left. -2\Delta\rho_{12}\Delta\rho_{23}h_1h_2h_3(h_1+h_2+h_3-\frac{h_1h_3}{h_2})\right]^{1/2}\,.
\end{eqnarray}

These two phase speeds are associated with symmetric and anti-symmetric oscillations of the interfaces,
\begin{eqnarray}
\eta_s (x,t) &=& \frac{1}{2}\left( \eta_{12}(x,t) + \eta_{23}(x,t)\right)\,,\\
\eta_a (x,t) &=& \frac{1}{2}\left( \eta_{12}(x,t) - \eta_{23}(x,t)\right)\,.
\end{eqnarray}
We will refer to mode-$s$ and mode-$a$ for $\eta_s$ and $\eta_a$ respectively.
We also define two Froude numbers $Fr_{s/a}=v/c_{s/a}^m$ which are the appropriate non-dimensional numbers to delimitate the oscillating regime as we will see in the following.
Numerical values from the experiments are summarized in Table~\ref{tab:eauxmortes_3couches_params}.

\begin{figure*}[!ht]
\begin{center}
\begin{picture}(17,7)
\put(0,0){\includegraphics[width=17\unitlength]{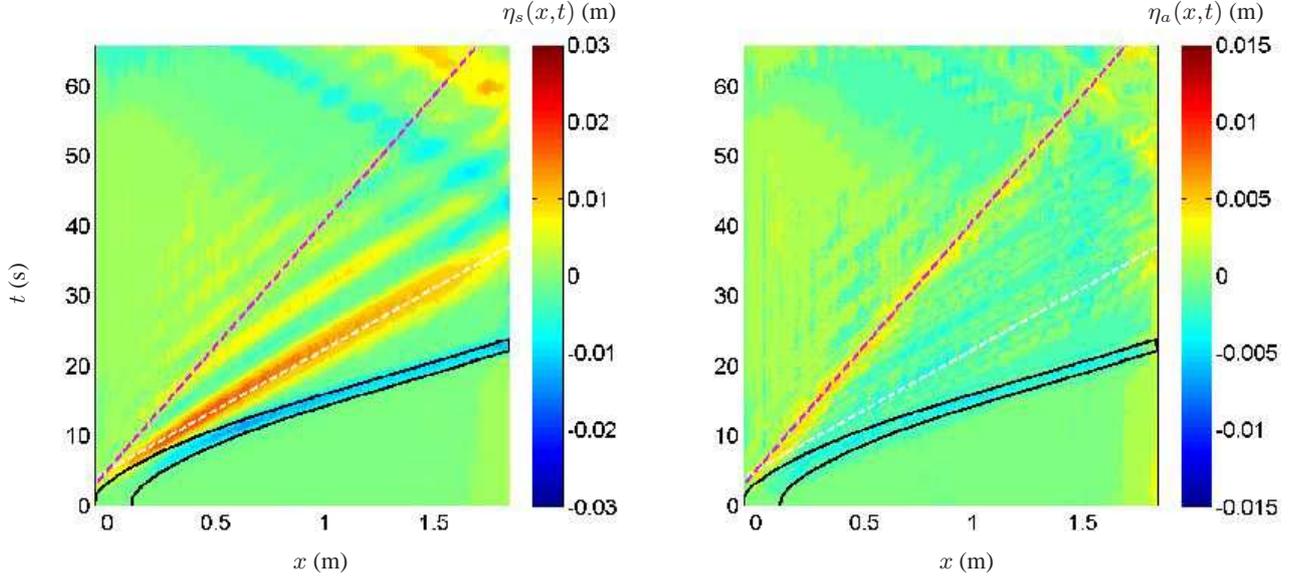}}
\put(4,-0.5){$x$ (m)}
\put(12.5,-0.5){$x$ (m)}
\put(0.25,3){\rotatebox{90}{$t$ (s)}}
\put(6.75,6.75){$\eta_{s}(x,t)$ (m)}
\put(15.25,6.75){$\eta_{a}(x,t)$ (m)}
\end{picture}
\end{center}
\caption{{\it Ballistic regime ($Fr_s>1$}). Spatio-temporal diagram of interfacial displacements $\eta_s(x,t)$ and $\eta_a(x,t)$ (in m). The solid black lines are the bow and stern of the boat. The magenta (resp. white) dotted line is an indication of the speed of propagation of $0.028$~m~s$^{-1}$ (resp. $0.0575$~m~s$^{-1}$). Experimental parameters are given in Table~\ref{tab:eauxmortes_3couches_params} and $S_b=24$~cm$^2$, $F_t=23.5$~mN.}\label{fig:Frs_O1_reflabo}
\end{figure*}

To emphasize that the two interfaces cannot be seen as independent oscillatory structures, since they are coupled by the intermediate layer, we give in 
table~\ref{tab:eauxmortes_3couches_params} the values of the Froude numbers for the two-layer equivalent solutions, $Fr_{12}$ and $Fr_{23}$. The values associated with the experiments do not allow the sub/supercritical classification.

\subsection{Case $Fr_s>1$ and $Fr_a>1$}

We first consider the case where the vessel is going faster than the fastest wave.
The experimental parameters associated with the results shown in Fig.~\ref{fig:Frs_O1_reflabo} are $S_b=24$~cm$^2$ and $F_t=23.5$~mN. The limit speed of the boat is $v_\ell\simeq0.11$~m~s$^{-1}$, which gives $Fr_s=1.5$ and $Fr_a=2.4$.
We only present the interfacial displacement as $\eta_s$ and $\eta_a$ because the spatio-temporal diagrams for $\eta_{12}$ and $\eta_{23}$ are very similar and do not clearly exhibit the link between the waves and the boat dynamics.

Both modes are generated initially but the amplitude of mode-$s$ is more than two times larger than that of mode-$a$ which is at the limit of detection.
Symmetric oscillations form a clear wave train whose front propagates at a speed of $0.0575$~m~s$^{-1}$, while showing a similar spreading of the wave crests as in the two-layer case: this is an indication of dispersion effects. Speed of propagation of mode-$a$ is $0.028$~m~s$^{-1}$.

\begin{figure*}[!ht]
\begin{center}
\begin{picture}(18,7)
\put(0,0){\includegraphics[width=17\unitlength]{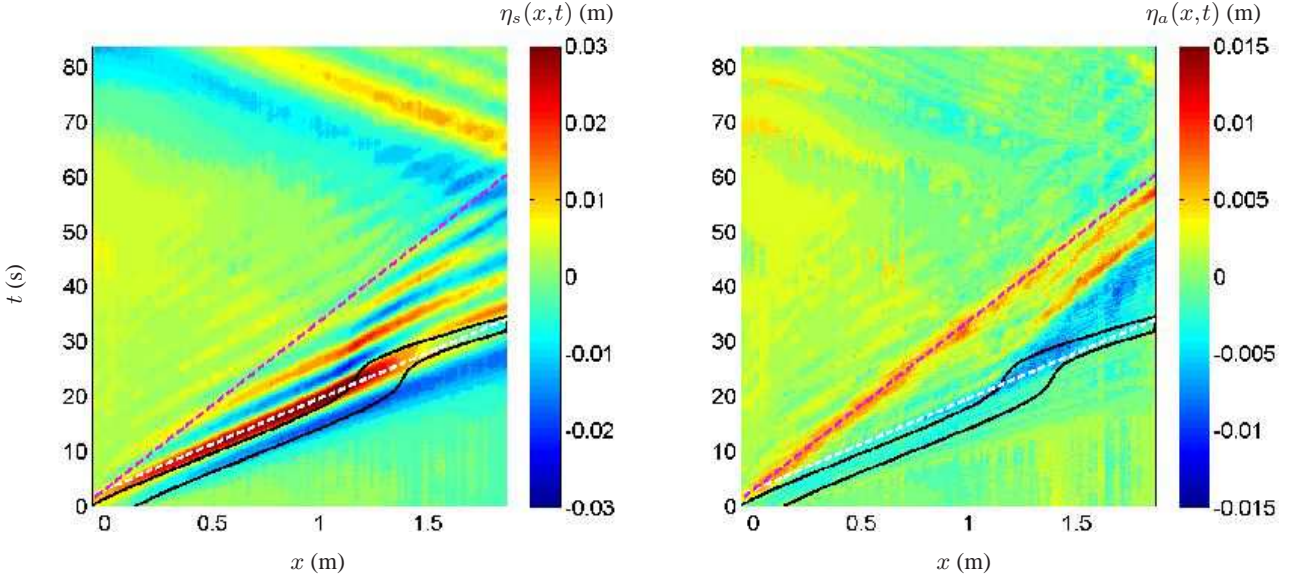}}
\put(4,-0.5){$x$ (m)}
\put(12.5,-0.5){$x$ (m)}
\put(0.25,3){\rotatebox{90}{$t$ (s)}}
\put(6.75,6.75){$\eta_{s}(x,t)$ (m)}
\put(15.25,6.75){$\eta_{a}(x,t)$ (m)}
\end{picture}
\end{center}
\caption{{\it Oscillating regime ($Fr_s<1$}).  Spatio-temporal diagram of interfacial displacements $\eta_s(x,t)$ and $\eta_a(x,t)$ (in m). The solid black lines are the bow and stern of the boat. The magenta (resp. white) dotted line is an indication of the speed of propagation of $0.0325$~m~s$^{-1}$ (resp. $0.06$~m~s$^{-1}$). Experimental parameters are given in in Table~\ref{tab:eauxmortes_3couches_params} and $S_b=24$~cm$^2$, $F_t=20.6$~mN.}\label{fig:Frs_o1_reflabo}
\end{figure*}

\subsection{Case $Fr_s<1$ and $Fr_a \simeq 1$}

We now consider a subcritical case. It is however difficult to define clearly this domain with respect to both modes. The experimental parameters associated with the results shown in Fig.~\ref{fig:Frs_o1_reflabo} are $S_b=24$~cm$^2$ and $F_t=20.6$~mN. The mean speed of the boat is $<v>\simeq0.06$~m~s$^{-1}$ which gives $Fr_s=0.6$ and $Fr_a=0.9$, with oscillations of the order of $0.04$~m~s$^{-1}$.

As observed previously, both modes are generated when the boat starts. Mode-$s$ remains two times larger than mode-$a$ but they are of noticeable amplitude.
The symmetric mode evolves with the boat and reproduces the amplification and steepening processes observed in the two-layer case. The mode-$s$ breaks on the boat when its amplitude is the largest, and a depression (symmetric for both interfaces) is expelled at the bow.
The characteristics of the mode-$s$, wavelength $\lambda_s\simeq0.4$~m and group velocity $c_{g,s}\simeq0.06$~m~s$^{-1}$ lead to an oscillating frequency of $0.15$~Hz, too slow for this visualization.

Concerning mode-$a$, its spatial structure is close to a solitary wave train, with a group velocity $c_{g,a}\simeq0.0325$~m~s$^{-1}$, close to $c_a^m$ (being $0.044$~m~s$^{-1}$). Each time it is generated, a strong acceleration of the boat occurs.

\subsection{Discussion}
\label{sec:discussion_3layer}
The unsteady behavior associated with dead-water is still observed in the three-layer fluid, with strong analogy to the two-layer case.
The stratification considered being more complex, we must consider two baroclinic modes associated with symmetric and anti-symmetric oscillations of the interfaces and that are also referred to as mode-$1$ and mode-$2$ in the literature.

\cite{bib:RusasGrue02} present solutions of the nonlinear equations that have strong similarities with our observations. More specifically, the spatio-temporal diagram of mode-$a$ exhibit solitary waves of mode-$2$ with oscillatory short mode-$1$ waves superimposed. This is in agreement with close values of the experiments with the numerical calculations (Boussinesq limit, $h_1/h_3\simeq1$ and $h_1/h_2\simeq1.7$).

Perturbations generated by the boat give birth to both modes, especially when the acceleration of the boat is important.
This result is consistent with the study of \cite{bib:Nicolaou95}, verified experimentally by \cite{bib:Robey97}, stating that an accelerated object in a continuously stratified fluid, with a Brunt--V\"{a}is\"{a}l\"{a} frequency $N(z)$, excites a continuum of modes whose vertical profile $w(z)$ is described by
\begin{equation}
\frac{d^2 w}{dz^2}(z) + k_x^2\left(\frac{N^2(z)}{\omega^2}-1\right)w(z)=0\,,
\label{eq:modes}
\end{equation}
along with the boundary conditions for the vertical velocity to be zero at the top and bottom.

Finally, we have observed that the mode-$1$ is strongly coupled to the dynamics of the boat, which corresponds to the fastest wave propagating in this stratification. A weak mode-$2$ is associated to noticeable acceleration of the boat, but evolves freely from the other mode.

%%%%%%%%%%%%%%%%%%%%%%%%%%%%%%%%%%%%%%%%%%%%%%%%%
%%%%%%%%%%%% CONTINUOUS CASE %%%%%%%%%%%%%%%%%%%%
\section{Continuously stratified fluid with a pycnocline}
\label{pycnocline}

We have observed that several waves are generated when the boat evolves in a complex stratification. In the case of a linearly stratified fluid, an infinite number of modes can propagate.
We have actually considered the case of a linearly stratified fluid with a pycnocline (see Fig.~\ref{fig:eauxmortes_linear_profil}). Several reasons can be invoked. It keeps the stratification undisrupted when the boat evolves in the top layer, it allows larger vertical displacements at the density jump leading to larger amplitudes, it can be modeled as a coupling between interfacial and internal waves, and it corresponds to a more realistic setup in comparison with observations made in natural environment.

Experimental parameters associated with the experiments presented are in Table~\ref{tab:eauxmortes_Nz_params}. We use technique-$2$ in order to observe both interfacial oscillations and internal waves.
\begin{table*}
\begin{tabular}{llccl}
\multicolumn{2}{l}{Parameters} & symbols & values & units \\
\hline\\
\multirow{2}{*}{Fluid 1} & density & $\rho_1$ & $0.9978$ & g~cm$^{-3}$ \\
		& depth & $h_1$ & $5.0$ & cm\\ 
\\
\multirow{3}{*}{Fluid 2} & density & $\rho(z)$ & $1.007-1.028$ & g~cm$^{-3}$ \\
		& depth & $h_2$ & $20.0$ & cm\\ 
		& BV frequency & $N$ & $0.9$ & rad~s$^{-1}$\\ 
\\
\multirow{4}{*}{Interface $\eta_{12}$} & mean density & $\bar{\rho_{12}}=\frac{\rho_2+\rho_1}{2}$ & $1.0024$ & g~cm$^{-3}$\\
 & density jump & $\Delta\rho_{12}=\frac{\rho_2-\rho_1}{\bar{\rho_{12}}}$ & $0.0092$ & \\
 & maximum phase speed & $c_{\phi,12}^m=\sqrt{\Delta\rho_{12} g\frac{h_1 h_2}{h_1+h_2}}$ & $0.06$ & m~s$^{-1}$\\
 & Froude number & $Fr_{12}=\frac{U}{c_{\phi,12}^m}$ & $0.6 - 2.3$ & \\
\\
\multirow{2}{*}{Modes} & maximum phase speed & $\frac{Nh}{\pi}$ & $0.072$ & m~s$^{-1}$\\
 & Froude number & $Fr_h=\frac{U\pi}{Nh}$ & $0.5 - 1.9$ & \\
\end{tabular}
%% NOTES exp strat 2
% exp 13 U~0.14m/s
% exp 11 U~0.m/s
\caption{Experimental parameters used for the experiments with a linearly stratified fluid with a pycnocline of depth $h=h_1+h_2$.}\label{tab:eauxmortes_Nz_params}
\end{table*}

Similarly to the three-layer case, the complex stratification considered here allows different definition for the Froude number. We know from the previous study that the one associated with the internal waves and based on the phase speed of the first vertical mode, $Fr_h=U\pi/Nh$, is the adequate one although its value is actually not so different from the one obtained from the equivalent two-layer fluid.

\subsection{Modal decomposition}
As discussed in Sec.~\ref{sec:discussion_3layer}, we expect the non-stationary evolution of the boat to generate several modes. We anticipate here by presenting the expected mode structures associated with the stratification (Fig.~\ref{fig:eauxmortes_linear_profil}) and obtained from solving Eq.~(\ref{eq:modes}).
From the vertical structure of the vertical component of the velocity obtained from (\ref{eq:modes}), we can deduce the corresponding density gradients, since
\begin{equation}
\rho^\prime_n(z)=\mathi\rho^\star \frac{N(z)^2}{\omega}w_n(z)\,,
\end{equation}
then
\begin{equation}
\partial_x\rho^\prime_n(z)=-\rho^\star \frac{k_x N(z)^2}{\omega}w_n(z)\,,
\end{equation}
where $\rho^\star$ is a reference density $\omega$ the frequency of the mode and $k_x$ its horizontal wavenumber.

\begin{figure}
\begin{center}
\begin{picture}(10,3.25)
\put(0.35,0){\includegraphics[width=4\unitlength]{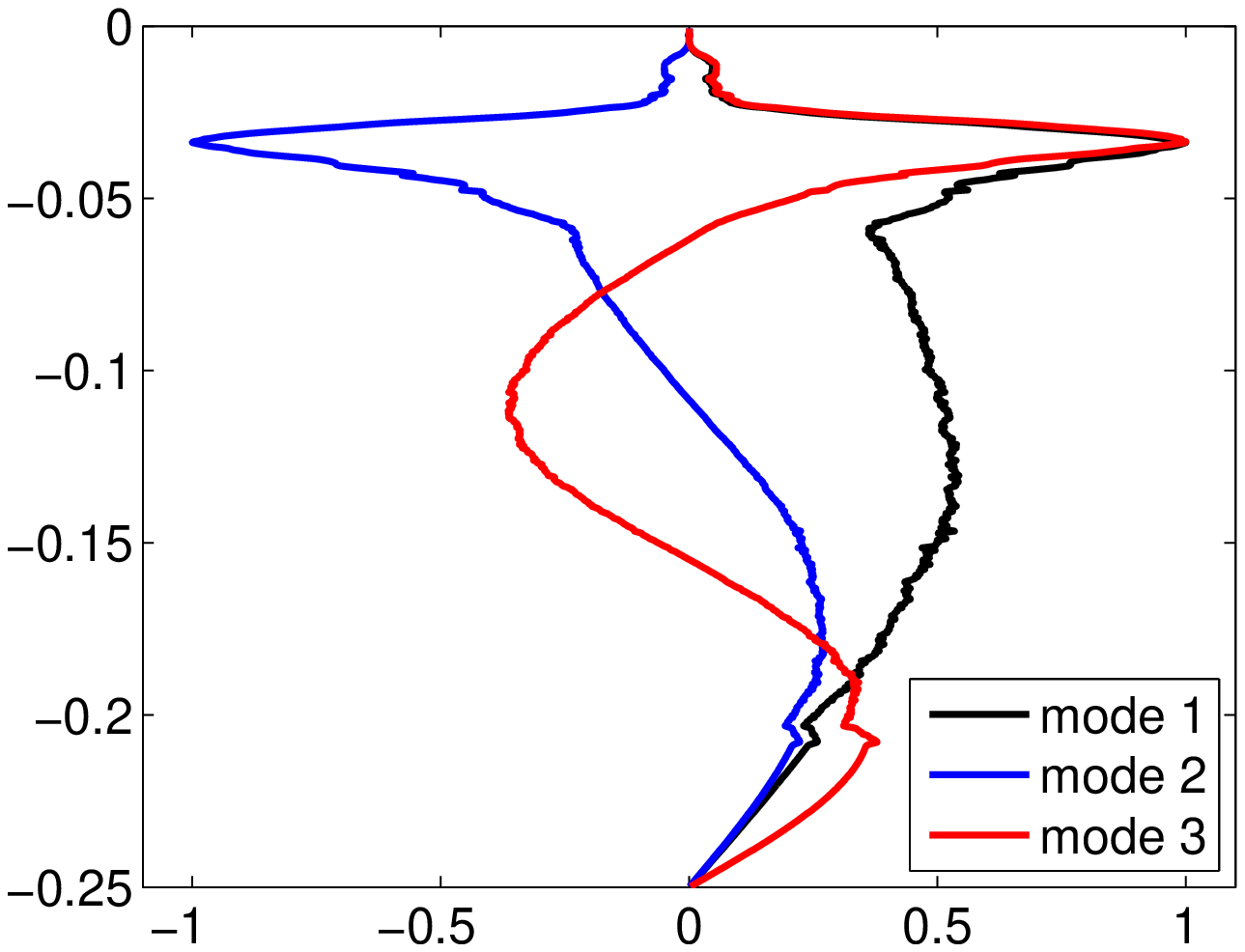}}
\put(4.45,0){\includegraphics[width=4\unitlength]{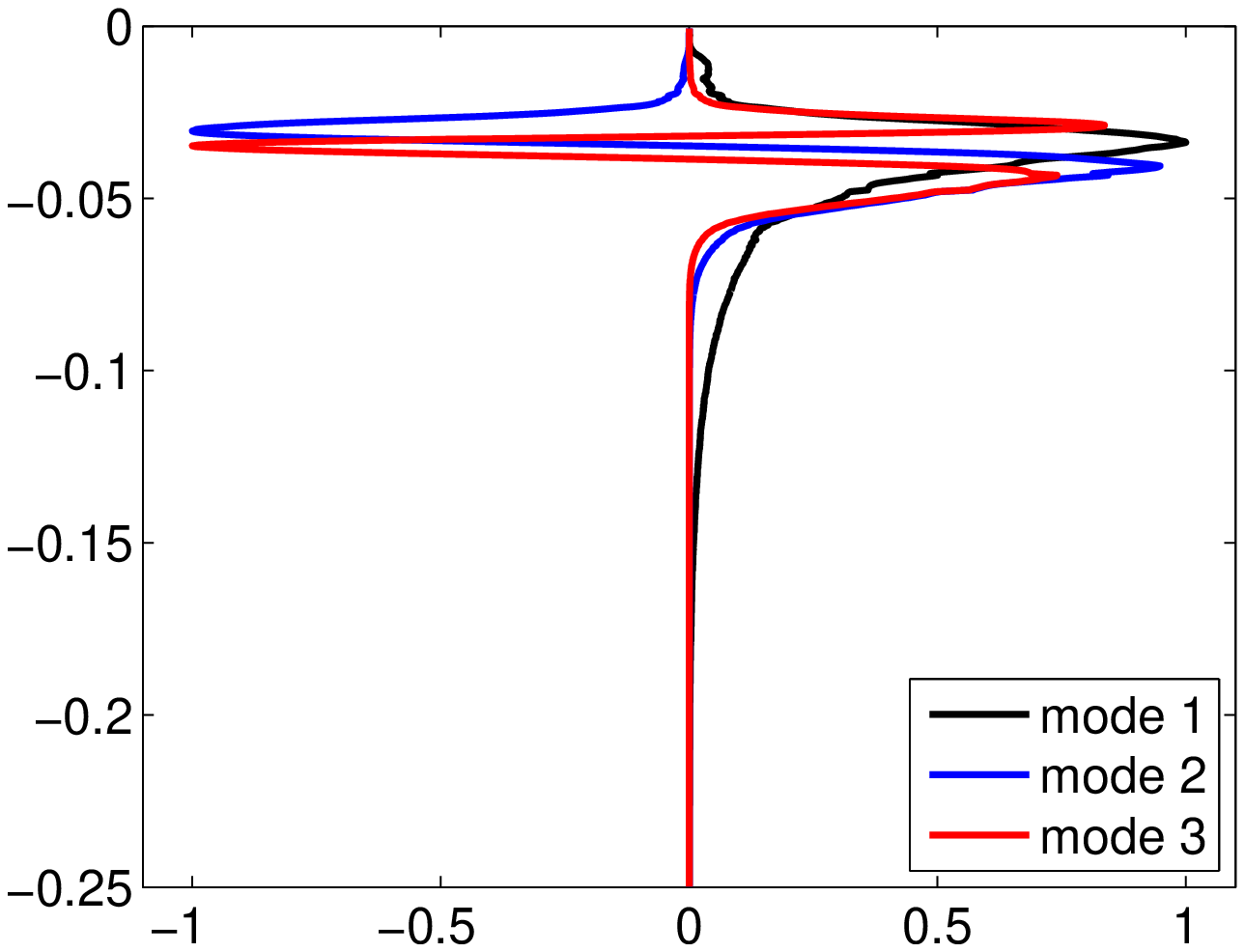}}
\put(2.4,3.15){(a)}
\put(6.5,3.15){(b)}
\put(0,1.3){\rotatebox{90}{$z$ (m)}}
\put(2.2,-0.5){$\partial_x\rho^\prime_n$}
\put(6.3,-0.5){$\partial_x\rho^\prime_n$}
\end{picture}
\end{center}
\caption{Normalized horizontal density gradient of the first three vertical modes of stratification in Fig.~\ref{fig:eauxmortes_linear_profil} with (a) $\omega_1=0.25$~rad~s$^{-1}$ and (b) $\omega_2=1.1$~rad~s$^{-1}$.}\label{fig:eauxmortes_Nz_modes}
\end{figure}

Figure~\ref{fig:eauxmortes_Nz_modes} gives the vertical profiles (of the horizontal density gradient) of the first three modes $\rho_n$ for two specific frequencies, $\omega_1=0.25$~rad~s$^{-1}$ in (a) and $\omega_2=1.1$~rad~s$^{-1}$ in (b). The former corresponds to a mode propagating in the total depth of the fluid (since $\omega_1<N$) whereas the latter is trapped in the pycnocline. ($\omega_2\in[0.9;1.8]$~rad~s$^{-1}$).

The modal decomposition depends on the frequency of study. In our case, the dead-water phenomenon is a non-periodic state and we expect perturbations that are not harmonic modes. However, from previous observations, it is reasonable to expect nonlinear evolution of modes evolving freely from the others. Based on this assumption, we expect the following projection to catch the key features of the dynamics.

The computed modes in Fig.~\ref{fig:eauxmortes_Nz_modes} will help us identify dominant structures associated with specific frequency bands, but it is important not to apply too selective time filtering to the data in order to conserve information about their complex dynamics.
Thus we extract the mode amplitudes $\hat{a}_n(x,t)$ after the two following steps done at each $x$-location:
\begin{enumerate}
\item a large band filtering centered around $\omega$ with a bandwidth $\Delta\omega$,
\item a projection of the vertical data onto the selected basis associated with the frequency band chosen $\omega$
\begin{equation}
\hat{a}_n(x,t)=\frac{1}{H}\int_{-H}^{0} \partial_x\rho^\prime_n(z).\partial_x\rho^\prime(x,z,t)\,dz \,.
\end{equation}
\end{enumerate}

\subsection{Case $Fr_h>1$}
The first experiment considered corresponds to the experimental parameters in Table~\ref{tab:eauxmortes_Nz_params} with $S_b=12$~cm$^2$ and $F_t=21.3$~mN.
The limit speed reached by the boat is $v_\ell=0.14$~m~s$^{-1}$ which corresponds to a supercritical case ($Fr_h=1.9$).

\begin{figure}
\begin{center}
\begin{picture}(10,7.5)
\put(0.4,3.75){\includegraphics[width=8\unitlength]{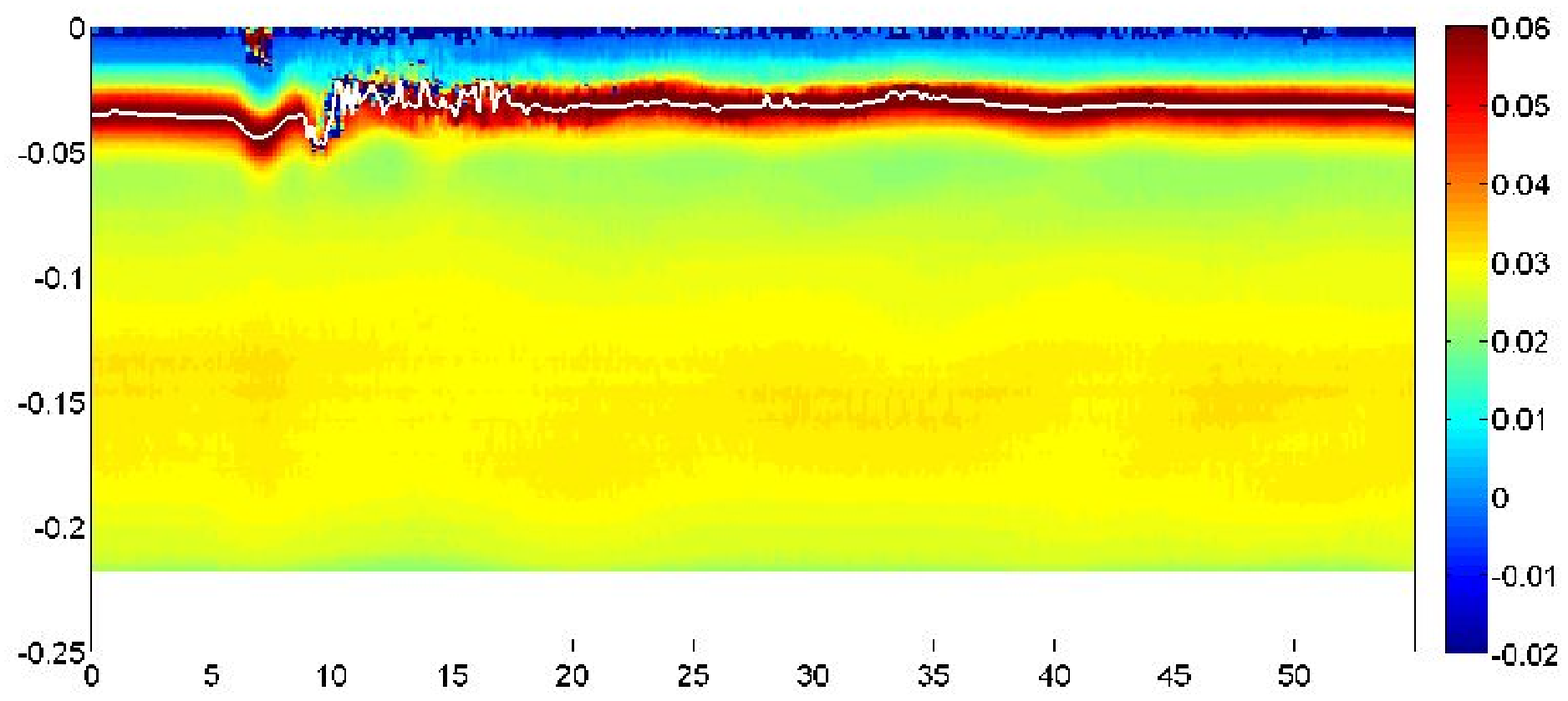}}
\put(0.4,0){\includegraphics[width=8\unitlength]{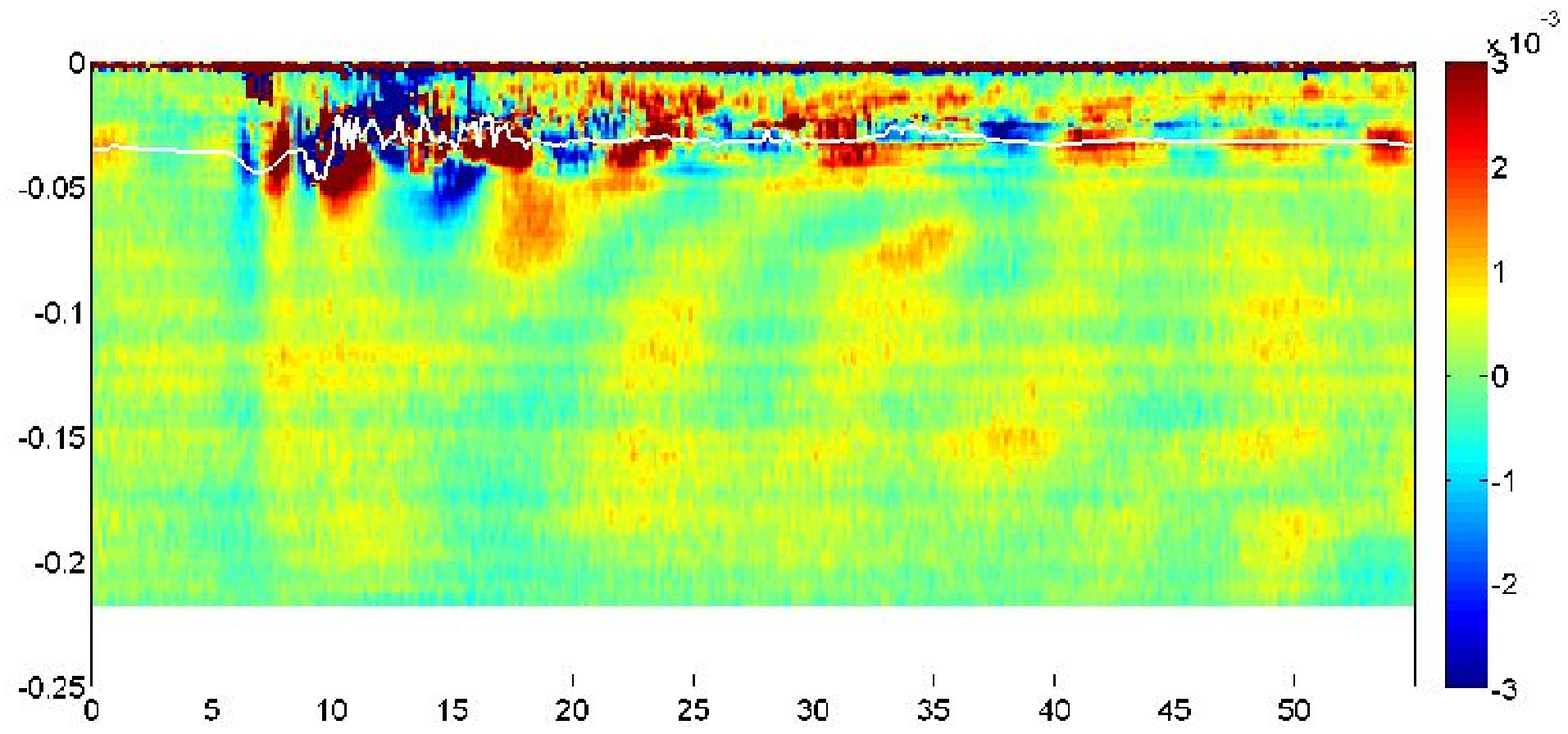}}
\put(0.5,3.5){(b)}
\put(0.5,7.25){(a)}
\put(4,-0.5){$t$ (s)}
\put(0,5){\rotatebox{90}{$z$ (m)}}
\put(0,1.25){\rotatebox{90}{$z$ (m)}}
\end{picture}
\end{center}
\caption{{\it $Fr_h>1$}. Time series of (a) $\partial_z\rho(x_0,z,t)$ and (b) $\partial_x\rho(x_0,z,t)$ extracted at $x_0=0.5$~m, in g/cm$^{4}$. }\label{fig:eauxmortes_Frh_01_drho}
\end{figure}

Figure~\ref{fig:eauxmortes_Frh_01_drho} presents time series of the density gradient  extracted on a vertical cut at $x_0=0.05$~m where the origin is taken at the bow of the boat at $t=0$. Interfacial displacements obtained from the evolution of the maximum of the vertical density gradient are given in white and are of less than $0.01$~m.
The fast evolution of the boat perturbs weakly the stratification and the signal associated with the internal waves field, for $z\in[-0.25,-0.05]$~m, is weak.

Two types of internal waves can still be discriminated.
Oscillations of the full water depth are triggered when the boat passes by at $t\simeq6$~s. Later on, some turbulent fluctuations in the pycnocline lead to small scales forcing radiating waves downwards ($t\in[10;20]$~s).
The modal projection leads to no clear signal whatever the frequency considered and are not given here.

\subsection{Case $Fr_h<1$}
The same configuration is used but with a force $F_t=11.0$~mN. The oscillatory dynamics of the boat around a mean speed $<v>\simeq0.036$~m~s$^{-1}$ corresponds to a subcritical case with $Fr_h=0.5$.

\begin{figure}
\begin{center}
\begin{picture}(10,7.5)
\put(0.4,3.75){\includegraphics[width=8\unitlength]{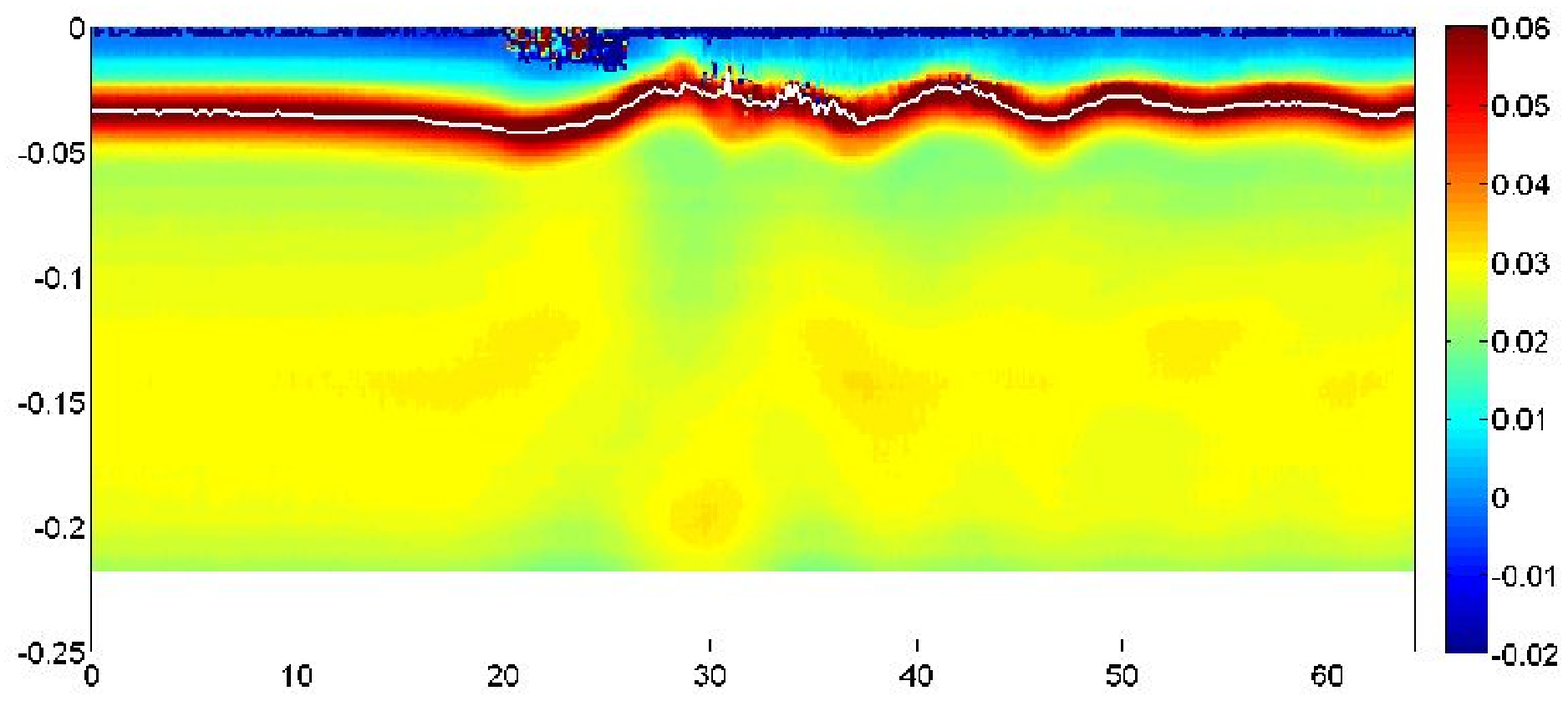}}
\put(0.4,0){\includegraphics[width=8\unitlength]{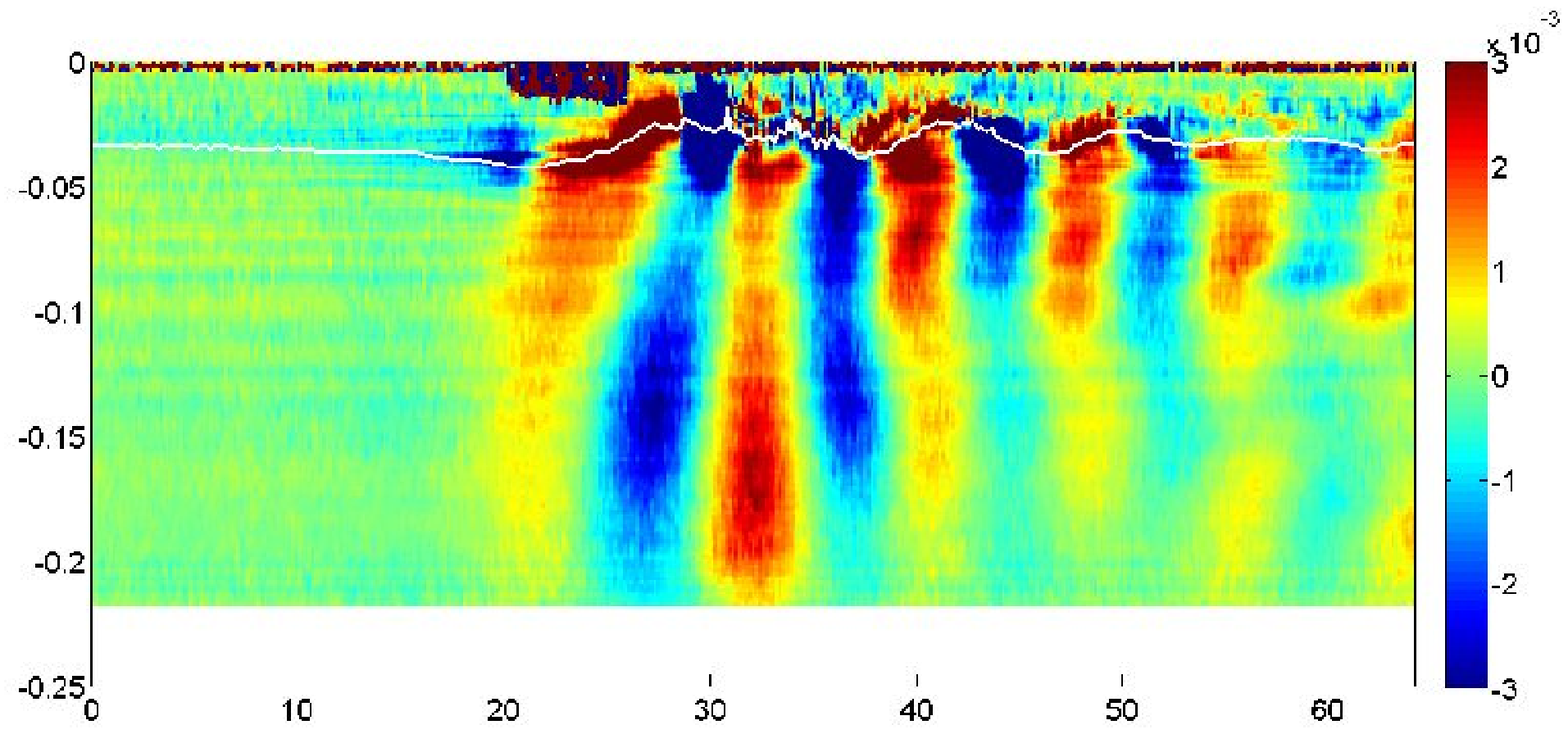}}
\put(0.5,3.5){(b)}
\put(0.5,7.25){(a)}
\put(4,-0.5){$t$ (s)}
\put(0,5){\rotatebox{90}{$z$ (m)}}
\put(0,1.25){\rotatebox{90}{$z$ (m)}}
\end{picture}
\end{center}
\caption{{\it $Fr_h<1$}. Time series of (a) $\partial_z\rho(x,z,t)$ and (b) $\partial_x\rho(x,z,t)$ extracted at $x=0.67$~m, in g/cm$^{4}$.}\label{fig:eauxmortes_Frh_o1_drho_x40}
\end{figure}

We represent the time series extracted at the location,  $x=0.67$~m (Fig.~\ref{fig:eauxmortes_Frh_o1_drho_x40}). The vertical cut corresponds to a region close to where the boat will stop and its speed is already decreasing. The stopping point of the boat is located at $x\simeq1.5$~m and occurs at $t\simeq50$~s.

In Fig.~\ref{fig:eauxmortes_Frh_o1_drho_x40}, the displacements of the pycnocline is of the order of $0.01$~m as in the supercritical case.
Nevertheless, the internal waves field is much more intense and the amplitude of the waves generated are larger as the boat evolves, suggesting the same nonlinear evolution of the waves as for interfacial waves.

\begin{figure}
\begin{center}
\begin{picture}(10,2.75)
\put(0.3,0){\includegraphics[height=2.39\unitlength]{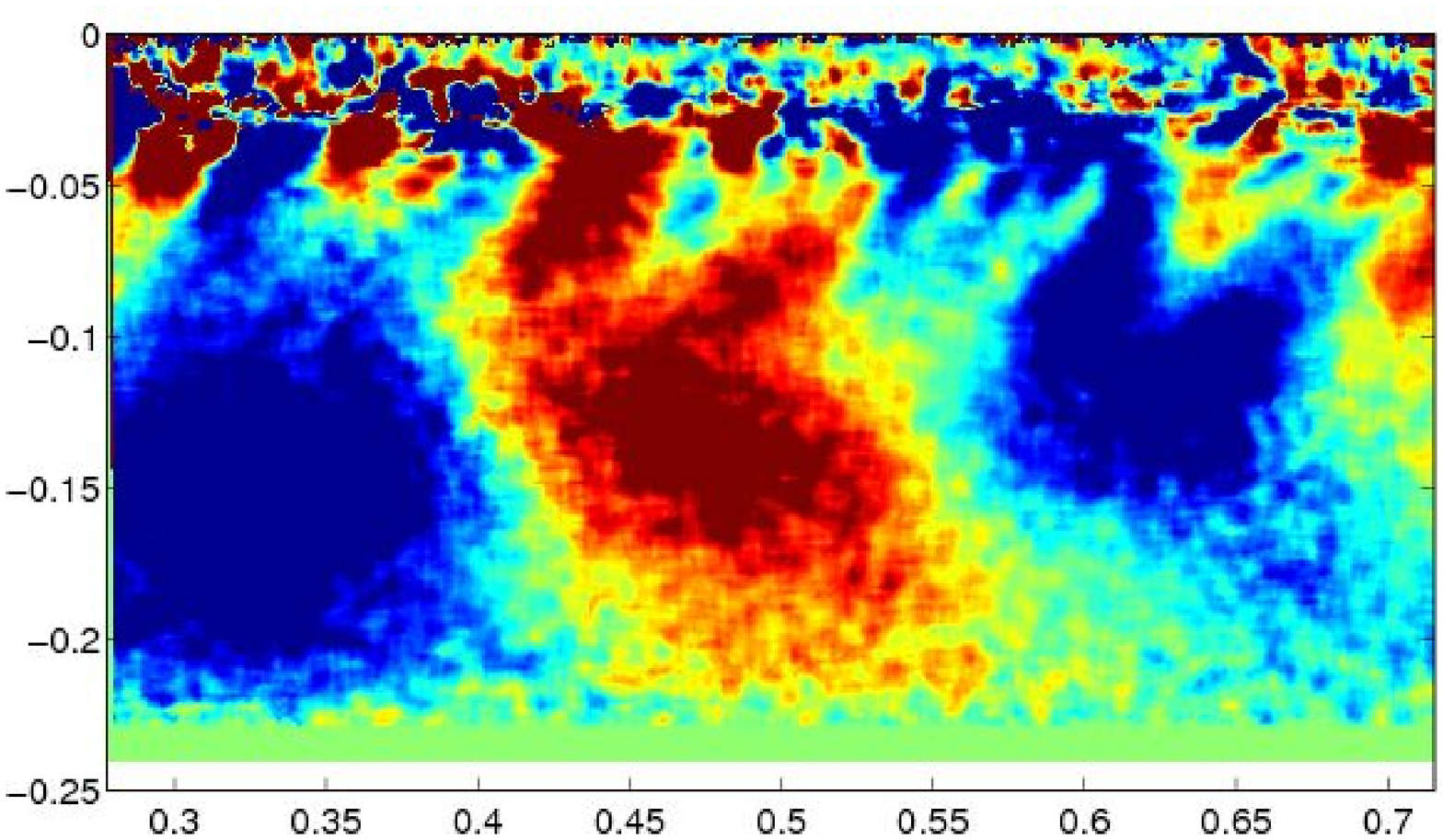}}
\put(4.5,0){\includegraphics[height=2.47\unitlength]{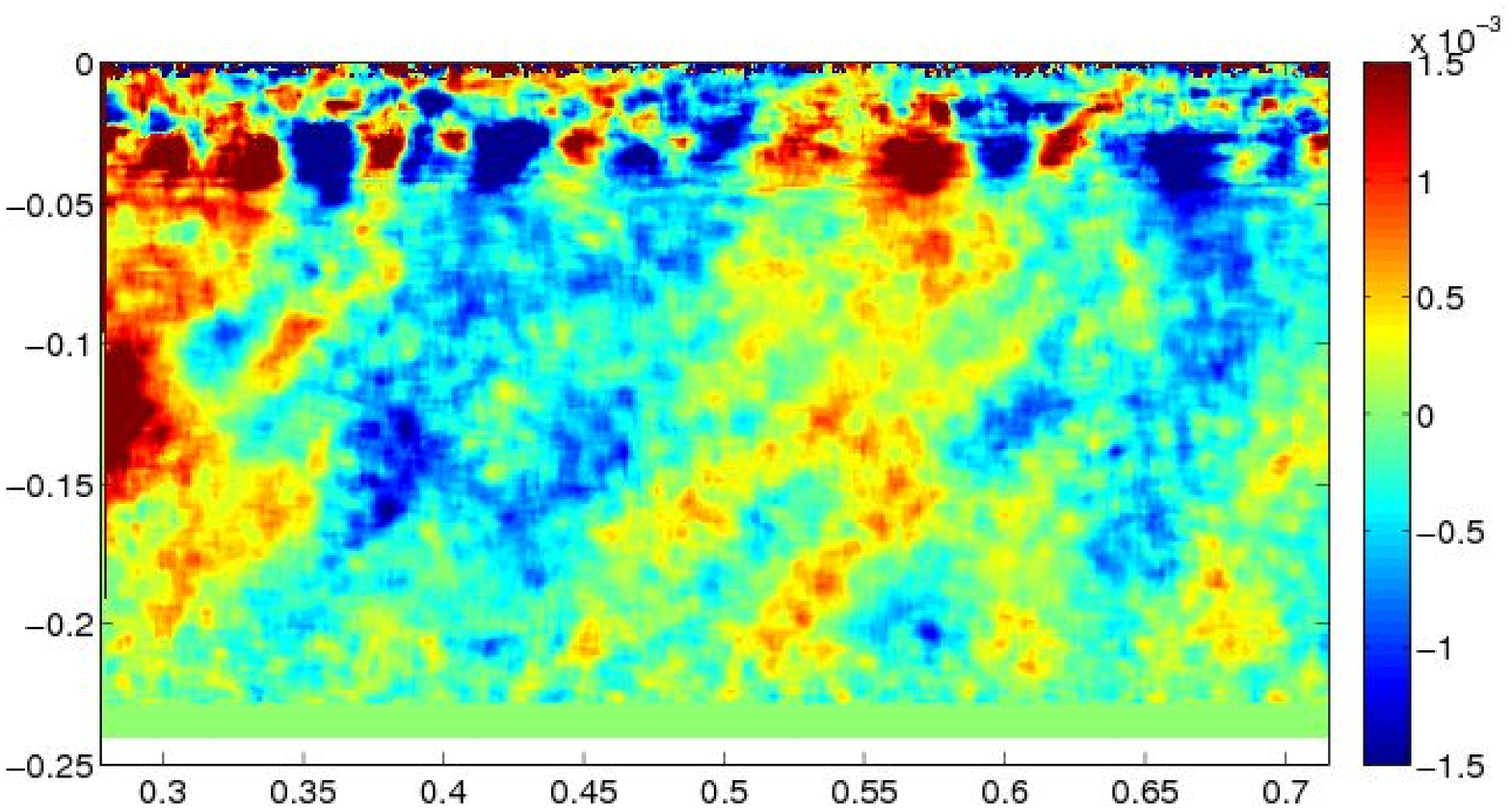}}
\put(2.2,2.5){(a)}
\put(6.4,2.5){(b)}
\put(2,-0.5){$x$ (m)}
\put(6.2,-0.5){$x$ (m)}
\put(0,1){\rotatebox{90}{$z$ (m)}}
\end{picture}
\end{center}
\caption{{\it $Fr_h<1$}. Instantaneous horizontal density gradients $\partial_x\rho(x,z,t_0)$ at times (a) $t_0=32$~s and (b) $t_0=61$~s, in g/cm$^{4}$.}\label{fig:eauxmortes_Frh_o1_dxrho}
\end{figure}

The spatial structure seems to be dominated by a mode-$1$ shape close to the boat, which is even more obvious in instantaneous visualizations of the density gradients as shown in Fig.~\ref{fig:eauxmortes_Frh_o1_dxrho}~(a). Other modes are also present in the wave field but appear later in time and are hard to discriminate from the radiated waves generated by turbulent patches at the pycnocline (Fig.~\ref{fig:eauxmortes_Frh_o1_dxrho}~(b)).

\begin{figure}
\begin{center}
\begin{picture}(10,14.5)
\put(0,9.5){\includegraphics[width=8.5\unitlength]{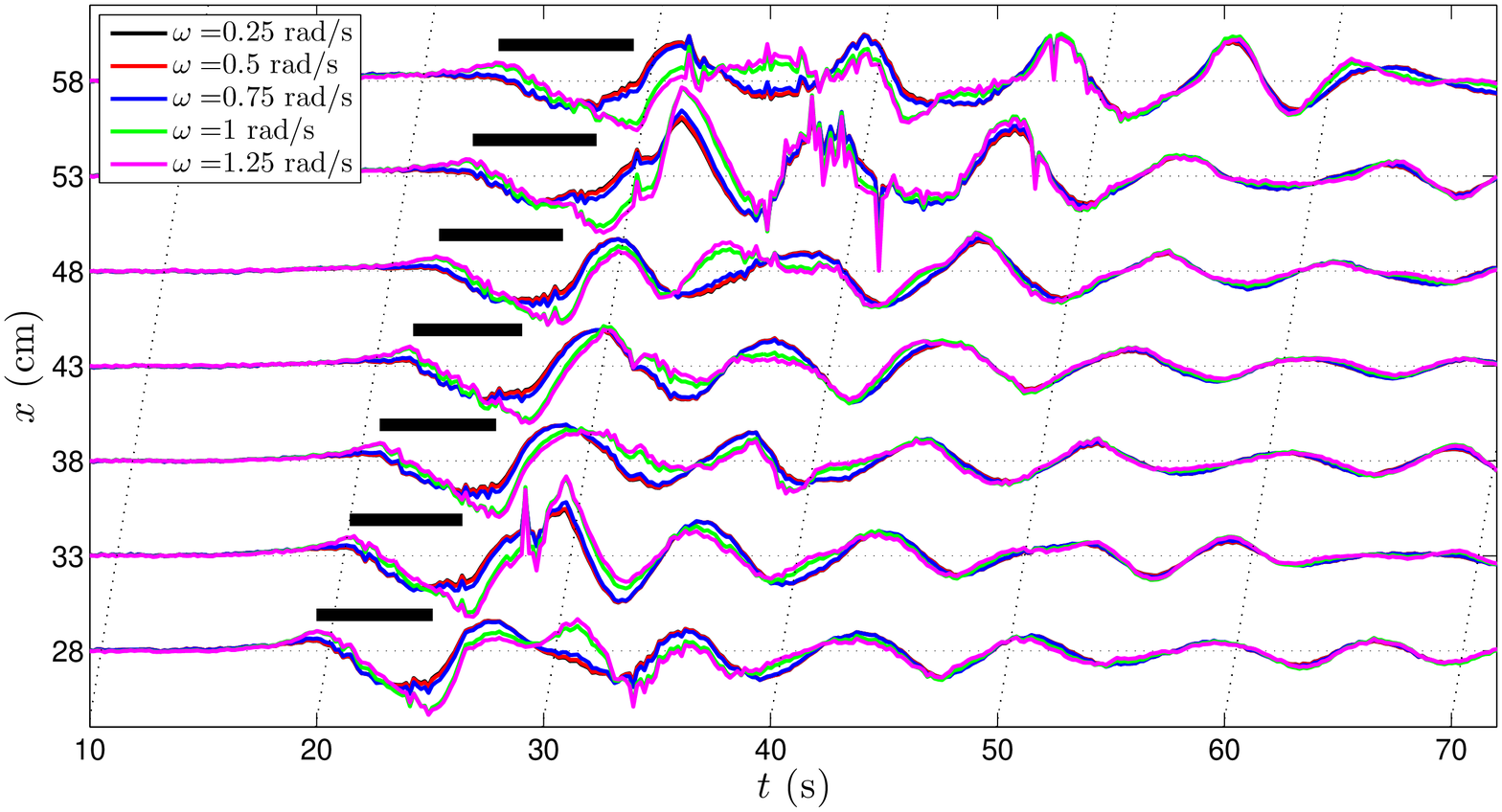}}
\put(4.5,14.25){(a)}
\put(0,4.5){\includegraphics[width=8.5\unitlength]{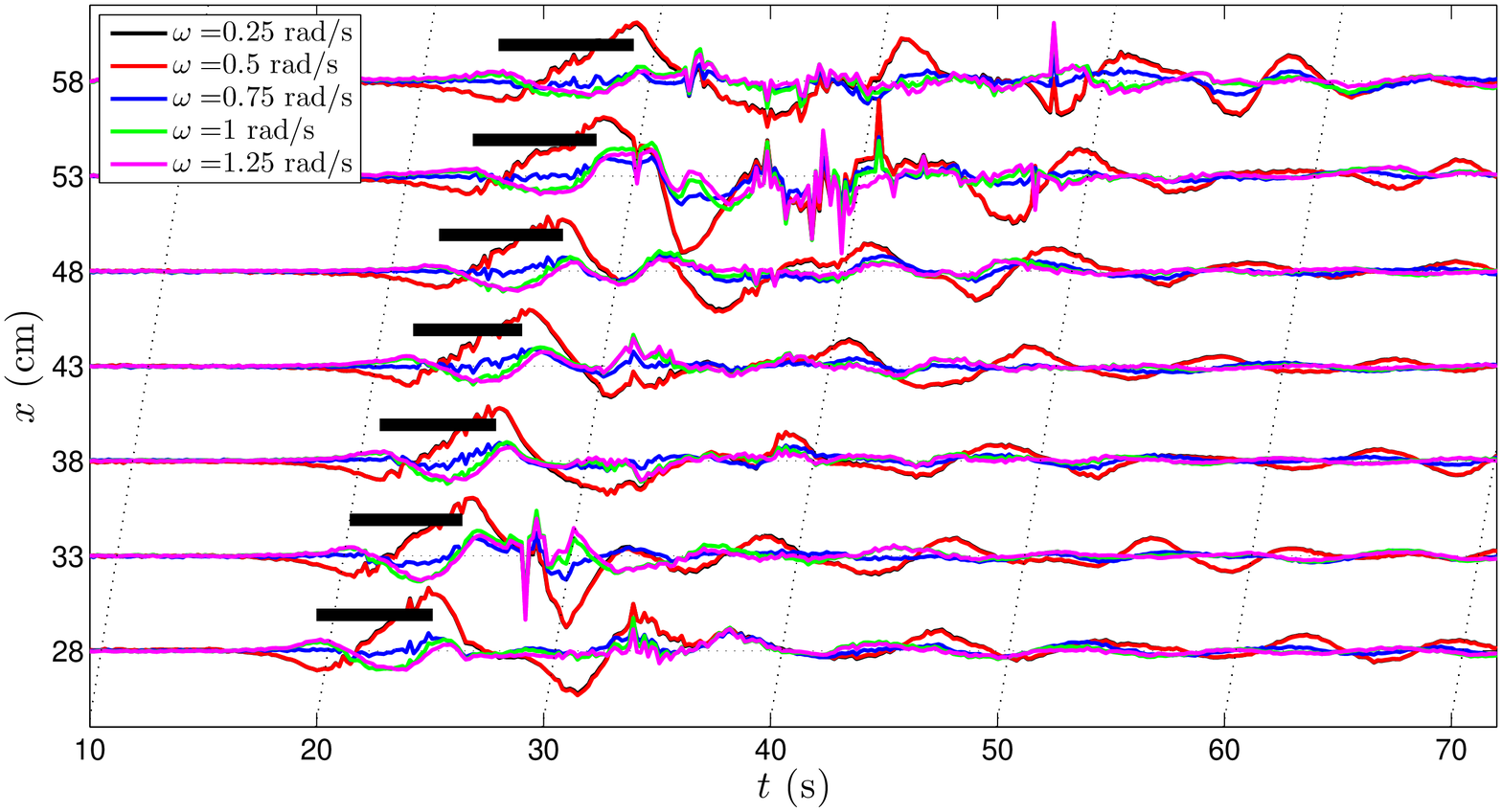}}
\put(4.5,9.25){(b)}
\put(0,-0.5){\includegraphics[width=8.5\unitlength]{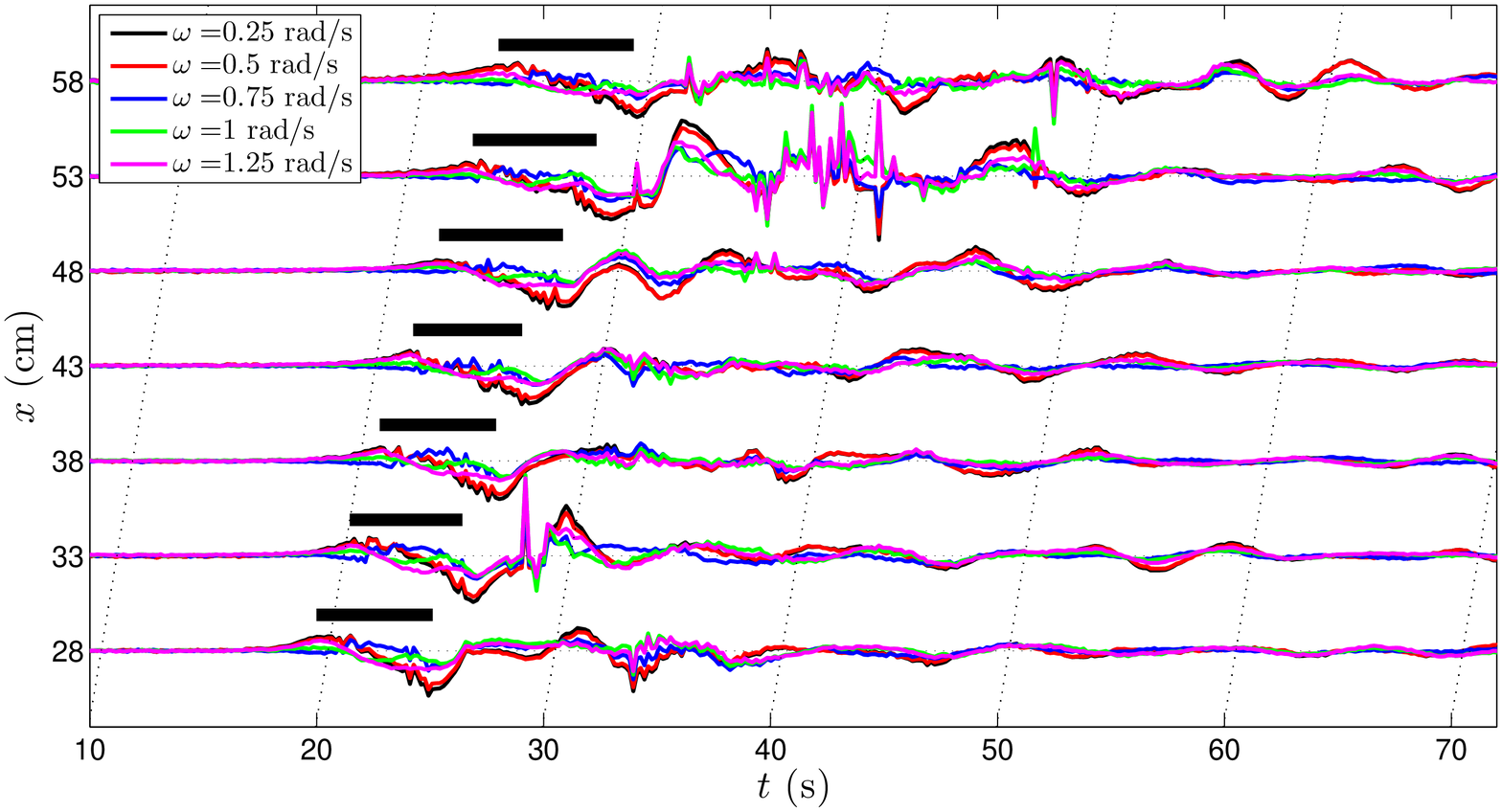}}
\put(4.5,4.25){(c)}
\end{picture}
\end{center}
\caption{{\it $Fr_h<1$}. Amplitudes $\hat{a}_n(x_0,t)$ with time of the projected modal structure at several $x_0$-locations for modes $n=1$ (a), $2$ (b) and $3$ (c) of the horizontal density gradient in arbitrary units. Different colors correspond to different frequencies associated with the modal basis. The tilted dashed lines corresponds to the characteristic 
speed of propagation $c_\phi^m=7.2$~cm/s. The black rectangle represents the boat passing through a vertical cross-section at each $x_0$-location studied.}\label{fig:Frh_o1_modes}
\end{figure}

In order to quantify in more detail the amplitude of each generated mode, we project the time series of the vertical structures extracted at different $x$-locations on the modal basis described before and computed at five different frequencies, $\omega=[0.25, 0.50, 0.75, 1.0, 1.25]$~rad~s$^{-1}$.
Large band filtering in time is done, with $\Delta\omega=0.25$~rad~s$^{-1}$, so as to keep track of the nonlinear evolution of the amplitude of the modes with time.
One can observe in Fig.~\ref{fig:Frh_o1_modes}~(a) that the amplitude of the mode-$1$ structure is indeed dominant since it is emitted for all frequencies considered, whereas the mode-$2$ in Fig.~\ref{fig:Frh_o1_modes}~(b) is of comparable amplitude at the lowest frequencies ($\omega=0.25$ and $0.50$~rad~s$^{-1}$) but is absent for larger values of $\omega$. Higher modes ($n\geq3$) are even weaker in amplitudes and only present at low frequencies (Fig.~\ref{fig:Frh_o1_modes}~(c)).

We notice that the initial perturbations for these low modes are generated below the boat since the black rectangles in Fig.~\ref{fig:Frh_o1_modes}~(a) and (b) are associated with a ramping amplitude of modes $1$ and $2$. The longer the rectangle is, the smaller the speed of the boat is, leading to larger initial perturbations which are very similar to what has been observed in the two-layer case (see Fig.~\ref{fig:eauxmortes_raidissement}).

Furthermore, the apparently cno\"idal (sharp crests and flat troughs) oscillations associated with the mode-$1$ internal waves observed in Fig.~\ref{fig:Frh_o1_modes}~(a) as the modes propagate, are a signature of the nonlinear dynamics of the internal waves.
Finally, by giving in all images the maximum phase speed of the waves with the tilted lines in Fig.~\ref{fig:Frh_o1_modes}, we can verify that the waves crests propagate at a smaller speed than $7.2$~cm~s$^{-1}$ but it is difficult to extract a constant speed with propagation for each mode.

\subsection{Discussion}
The dead-water phenomenon has been observed in the more general case of a linearly stratified fluid with a pycnocline.
The complex dynamics of the boat is coupled to the first mode of the stratification with frequencies below the maximum value of $N(z)$.

Although the wave field associated with the subcritical regime is analogous to the previous layered stratifications considered, the supercritical regime is different in nature since it consists mainly of radiated waves from turbulent perturbations in the pycnocline. This is similar to observations of the radiated wave field associated with an object moving at constant speed in a stratified fluid~\citep{bib:RottmanetalAPS2004}.

\conclusions
\label{sct:conclusion}

By revisiting the historical experiments of Ekman's PhD Thesis and extending it to more general stratifications, we have shown the robustness of the dead-water phenomenon.
The experimental techniques used have revealed new insights on the century-old problem.

One important characteristic of the dead-water phenomenon is that the dynamics of the boat is coupled to the fastest mode of the stratification considered, although several modes are generated at each acceleration of the boat.
Furthermore, the nonlinear features of the phenomenon must be considered in order to describe analytically its unsteady nature. Classical models such as presented by \citeauthor{bib:Ekman1904} or \citeauthor{bib:Milohetal93} are not sufficient.

It would be of great interest to provide an analytical description of the coupled dynamics of the boat and the waves it generates in order to take into account unsteady behaviors.

An extended study of the problem in a three-dimensional setup is in preparation but we are still developing a collaboration with Playmobil\,\copyright.

\begin{acknowledgements}
The authors are very thankful to Leo Maas for introducing the topic of dead-water to them, for many discussions and for his great knowledge of the historical background. TD thanks Peter Morgan and Christina Kraus for mentioning similar reports in the Latin literature.
\end{acknowledgements}

\bibliographystyle{copernicus}
\bibliography{biblio_IW}
\end{document}